\documentclass{ametsocV6.1}

\title{The impact of transience in the interaction between orographic gravity waves and mean flow}

\authors{\correspondingauthor{Felix Jochum, jochum@iau.uni-frankfurt.de}Felix Jochum,\aff{a} Ray Chew,\aff{a} Fran\c{c}ois Lott,\aff{b} Georg S. Voelker\thanks{Georg S. Voelker's current affiliation: Leibniz-Institut f\"{u}r Ostseeforschung Warnem\"{u}nde, Rostock, Germany},\aff{a} Jan Weinkaemmerer\thanks{Jan Weinkaemmerer's current affiliation: Institut f\"{u}r Bio- und Geowissenschaften: Agrosph\"{a}re (IBG-3), Forschungszentrum J\"{u}lich GmbH, J\"{u}lich, Germany},\aff{a} Ulrich Achatz\aff{a}}

\affiliation{\aff{a}{Institut f\"{u}r Atmosph\"{a}re und Umwelt, Goethe-Universit\"{a}t Frankfurt, Frankfurt am Main, Germany}\\
\aff{b}{Laboratoire de M\'{e}t\'{e}orologie Dynamique, \'{E}cole Normale Sup\'{e}rieure, Paris, France}}

\abstract{A Lagrangian gravity-wave parameterization (MS-GWaM, Multi-Scale Gravity-Wave Model) that allows for fully transient wave-mean-flow interaction and horizontal propagation is applied to orographic gravity waves for the first time. Both linear and nonlinear mountain waves are modeled in idealized simulations within the pseudo-incompressible flow solver PincFlow. Two-dimensional flows over monochromatic orographies are considered, using MS-GWaM either in its fully transient implementation or in a steady-state implementation that represents classic mountain-wave parameterizations. Comparisons of wave-resolving simulations (not using MS-GWaM) and coarse-resolution simulations (using MS-GWaM) show that allowing for transience leads to a significantly more accurate forcing of the resolved mean flow. The model is able to reproduce the transient forcing of linearly generated mountain waves that slowly propagate upwards, in contrast to the instantaneous distribution of wave energy in classic parameterizations. At high altitudes, wave breaking induces a wind reversal that is captured by the transient model but inhibited in steady-state simulations, due to the assumption of critical level formation. This shows that transience can have a substantial impact in the interaction between mountain waves and mean flow.}

\begin{document}

\begin{center}
  \textbf{This work has been submitted to the Journal of the Atmospheric Sciences. Copyright in this work may be transferred without further notice.}
\end{center}

\maketitle

\newcommand{\ddd}{\mathrm{d}}
\newcommand{\mdt}[1]{\frac{\mathrm{D} #1}{\mathrm{D} t}}
\newcommand{\ddt}[1]{\frac{\mathrm{d} #1}{\mathrm{d} t}}
\newcommand{\ddx}[1]{\frac{\mathrm{d} #1}{\mathrm{d} x}}
\newcommand{\ddy}[1]{\frac{\mathrm{d} #1}{\mathrm{d} y}}
\newcommand{\ddz}[1]{\frac{\mathrm{d} #1}{\mathrm{d} z}}
\newcommand{\pdt}[1]{\frac{\partial #1}{\partial t}}
\newcommand{\pdx}[1]{\frac{\partial #1}{\partial x}}
\newcommand{\pdy}[1]{\frac{\partial #1}{\partial y}}
\newcommand{\pdz}[1]{\frac{\partial #1}{\partial z}}

\newcommand{\sgn}{\mathrm{sgn}}

\renewcommand{\div}{\mathbf{\nabla} \cdot}
\newcommand{\gra}{\mathbf{\nabla}}
\newcommand{\rot}{\mathbf{\nabla} \times}
\newcommand{\lap}{\mathbf{\nabla}^2}

\newcommand{\ros}{\mathrm{Ro}}
\newcommand{\mac}{\mathrm{Ma}}
\newcommand{\fro}{\mathrm{Fr}}
\newcommand{\rey}{\mathrm{Re}}
\newcommand{\lon}{\mathrm{Lo}}

\renewcommand{\Re}{\mathrm{Re}}
\renewcommand{\Im}{\mathrm{Im}}

\newcommand{\lef}{\left}
\newcommand{\rig}{\right}
\renewcommand{\vec}[1]{\mathbf{#1}}
\renewcommand{\bar}[1]{{\overline{#1}}}
\renewcommand{\hat}[1]{{\widehat{#1}}}
\newcommand{\til}[1]{{\widetilde{#1}}}

\section{Introduction}
\label{section_introduction}

  Gravity waves are well-known to significantly impact atmospheric dynamics. Generated by sources such as convection, flow over orography and jet imbalances, they propagate upwards and thus transport momentum and energy to high altitudes. Atmospheric models used in numerical weather prediction and climate projections typically cannot fully resolve the gravity-wave spectrum, so that parameterizations for their impact on the resolved flow are needed \citep{Fritts2003, Alexander2010, Achatz2024}.

  Most of these are currently based on single-column and steady-state approximations, under which temporal changes in gravity-wave sources affect momentum-flux distributions instantaneously and in the vertical direction only. Consequently, horizontal wave propagation and transient interactions with the mean flow are neglected \citep{Boeloeni2021}. As has been shown in various studies \citep[e.g.][]{Sato2009, Senf2011, Ehard2017, Muraschko2015, Boeloeni2016, Wei2019}, these simplifications can lead to significant errors in the prediction of the resolved flow, encouraging the use of models with fewer limitations.

  MS-GWaM is a multi-scale gravity-wave model based on WKB theory. Developed initially by \citet{Muraschko2015}, \citet{Boeloeni2016}, \citet{Wilhelm2018} and \citet{Wei2019}, it uses ray-tracing techniques to solve prognostic equations for the properties of locally monochromatic gravity-wave fields and computes the corresponding mean-flow impact. One version of it is interactively coupled to PincFlow, a pseudo-incompressible flow solver developed by \citet{Rieper2013} and \citet{Schmid2021}. This coupling is used in the present work to test the implementation of an orographic source, which does not yet exist in other versions of the model \citep[e.g.][]{Boeloeni2021, Kim2021, Kim2024, Voelker2024}. A steady-state mode is implemented to represent classic gravity-wave parameterizations.

  The tests are conducted in an idealized setting that features a two-dimensional, isothermal atmosphere with a monochromatic orography as a lower boundary. Multiple sets of WKB simulations are run with mountain heights ranging from the linear (low mountains) to the nonlinear regime (high mountains). The results of corresponding wave-resolving simulations are used as a reference.

  The following section gives an overview of the theoretical concepts on which MS-GWaM is built, as well as a description of the model components that are of particular significance in the present context. Section \ref{section_results} discusses the results obtained in the idealized simulations. Lastly, section \ref{section_conclusion} gives a summary and conclusion.

\section{Theory}
\label{section_theory}

  MS-GWaM is based on a WKB theory built around a scale separation between the mean flow and gravity waves. The Euler equations form the basis from which the gravity-wave dispersion and polarization relations, and ultimately prognostic equations for the essential wave properties are inferred. A detailed derivation can be found in \citet{Achatz2017} and \citet{Achatz2023}. This section summarizes the results of the theory, before giving a description of the orographic source and illustrating how the equations can be reduced to the single-column steady-state limit. The appendixes contain additional information on the coordinate system (A), discretization (B) and ray-tracing methods (C-E) of the model.

  \subsection{Transient theory}
  \label{subsection_transient_theory}

    The gravity-wave dispersion relation may be written as $\omega \lef(\vec{x}, t\rig) = \Omega \lef[\vec{x}, t, \vec{k} \lef(\vec{x}, t\rig)\rig]$, with
    \begin{align}
      \label{dispersion_relation}
      \Omega \lef(\vec{x}, t, \vec{k}\rig) = \vec{k} \cdot \vec{u}_\mathrm{m} \lef(\vec{x}, t\rig) \pm \sqrt{\frac{N^2 \lef(z\rig) \lef(k^2 + l^2\rig)}{\lef|\vec{k}\rig|^2}},
    \end{align}
    where $\vec{x} = \lef(x, y, z\rig)^\mathrm{T}$, $t$ and $\vec{k} = \lef(k, l, m\rig)^\mathrm{T}$ are the position vector, time and wave vector, $\vec{u}_\mathrm{m} = \lef(u_\mathrm{m}, \upsilon_\mathrm{m}, 0\rig)^\mathrm{T}$ is the mean wind and $N$ the buoyancy frequency. The Coriolis force is neglected in the present work. By expressing the gravity-wave frequency in terms of $\Omega$, its spatio-temporal dependence is split into an explicit contribution from variations in $\vec{u}_\mathrm{m}$ and $N$, and an implicit contribution that describes the corresponding dependence of $\vec{k}$. This can be used to derive the eikonal equations
    \begin{align}
      \label{ray_equations_frequency}
      \lef(\frac{\partial}{\partial t} + \vec{c}_\mathrm{g} \cdot \gra\rig) \omega & = \frac{\partial \Omega}{\partial t},\\
      \label{ray_equations_wavenumbers}
      \lef(\frac{\partial}{\partial t} + \vec{c}_\mathrm{g} \cdot \gra\rig) \vec{k} & = - \gra \Omega,
    \end{align}
    where $\vec{c}_\mathrm{g} = \lef(\partial_k \Omega, \partial_l \Omega, \partial_m \Omega\rig)^\mathrm{T}$ is the group velocity. MS-GWaM uses a Lagrangian approach in which these equations are integrated along the paths of wave propagation, i.e. so-called rays defined by
    \begin{align}
      \label{ray_equations_position}
      \ddt{\vec{x}} & = \vec{c}_\mathrm{g}
    \end{align}
    \citep{Achatz2017, Achatz2023}.

    This locally monochromatic approach is generalized to gravity-wave spectra via superposition of quasilinear wave modes. The evolution of the total wave-action density $\mathcal{A} = \sum_\alpha \mathcal{A}_\alpha$ is then governed by the conservation equation
    \begin{align}
      \label{phase_space_wave_action_conservation}
      \pdt{\mathcal{N}} + \vec{c}_\mathrm{g} \cdot \gra \mathcal{N} + \dot{\vec{k}} \cdot \gra_{\vec{k}} \mathcal{N} = \sum\limits_s \mathcal{S}_s
    \end{align}
    for the phase-space wave-action density
    \begin{align}
      \label{phase_space_wave_action_density}
      \mathcal{N} \lef(\vec{x}, t, \vec{k}\rig) = \sum\limits_\alpha \mathcal{A}_\alpha \lef(\vec{x}, t\rig) \delta \lef[\vec{k} - \vec{k}_\alpha \lef(\vec{x}, t\rig)\rig],
    \end{align}
    where $\alpha$ is an index for the (possibly infinitely many) contributing spectral components and $\delta$ is the delta distribution. In \eqref{phase_space_wave_action_conservation}, $\dot{\vec{k}} = \lef(\partial_t + \vec{c}_\mathrm{g} \cdot \gra\rig) \vec{k}$ can be understood as a spectral group velocity, $\gra_\vec{k} = \lef(\partial_k, \partial_l, \partial_m\rig)^\mathrm{T}$ denotes the spectral gradient and $\mathcal{S}_s$ are sinks and sources. Thus, in the absence of the latter, $\mathcal{N}$ is conserved along rays defined by
    \begin{align}
      \label{phase_space_rays}
      \ddt{} \lef(\vec{x}, \vec{k}\rig) = \lef(\vec{c}_\mathrm{g}, \dot{\vec{k}}\rig).
    \end{align}
    Moreover, because the phase-space velocity $\lef(\vec{c}_\mathrm{g}, \dot{\vec{k}}\rig)$ is non-divergent, the integral of $\mathcal{N}$ over phase space is conserved. MS-GWaM makes use of these properties by integrating \eqref{phase_space_wave_action_conservation} in a Lagrangian frame defined by \eqref{phase_space_rays} \citep{Achatz2017, Achatz2023}.

    The theory is completed with relations between the tendencies of the large-scale flow and divergences of small-scale fluxes. In particular, the impact of gravity waves on the horizontal-momentum equation is described by
    \begin{align}
      \label{mean_flow_impact}
      \lef(\pdt{u_\mathrm{m}}, \pdt{\upsilon_\mathrm{m}}\rig)_\mathrm{w} & = - \frac{1}{\bar{\rho}} \div{\int \hat{\vec{c}}_\mathrm{g} \lef(k, l\rig) \mathcal{N} \, \ddd V_\vec{k}},
    \end{align}
    where $\bar{\rho}$ is the density of a hydrostatic reference atmosphere, $\hat{\vec{c}}_\mathrm{g} = \vec{c}_\mathrm{g} - \vec{u}_\mathrm{m}$ is the intrinsic group velocity and $\ddd V_\vec{k} = \ddd k \ddd l \ddd m$. Note that the integral in \eqref{mean_flow_impact} is the so-called pseudo-momentum flux, which is equal to the physical gravity-wave momentum flux in the absence of rotation \citep{Achatz2017, Achatz2023}.

  \subsection{Wave breaking}
  \label{subsection_wave_breaking}

    WKB theory does not describe the turbulent process of waves breaking at large amplitudes, so that a saturation scheme is needed to capture this. Saturation is assumed to occur when the static-instability criterion is locally fulfilled. Wave breaking then leads to the onset of turbulence with fluxes that are described by a turbulent viscosity and diffusivity $D$ in a flux-gradient ansatz \citep[see][]{Lindzen1981, Becker2004}. The ensuing damping of wave amplitudes is represented by the sink
    \begin{align}
      \mathcal{S}_0 = - 2 D \lef|\vec{k}\rig|^2 \mathcal{N}
    \end{align}
    in the phase-space-wave-action-density equation. By performing an explicit Euler integration over a time step $\Delta t$ and demanding the change to be such that the instability criterion is no longer satisfied, one finds that the turbulent viscosity and diffusivity may be written as
    \begin{align}
      \label{diffusion_coefficient}
      D = \frac{\bar{\rho}}{4 \Delta t} \lef[\int N^4 \lef(k^2 + l^2\rig) m^2 \frac{\mathcal{N}}{\hat{\omega}} \, \ddd V_\vec{k}\rig]^{- 1} \max \lef[0, \frac{2}{\bar{\rho}} \int \frac{N^4 \lef(k^2 + l^2\rig) m^2}{\hat{\omega} \lef|\vec{k}\rig|^2} \mathcal{N} \, \ddd V_\vec{k} - \alpha_\mathrm{d}^2 N^4\rig],
    \end{align}
    where $\hat{\omega} = \omega - \vec{k} \cdot \vec{u}_\mathrm{m}$ is the intrinsic frequency and $\alpha_\mathrm{d}$ is a saturation coefficient that accounts for uncertainties of the criterion \citep{Boeloeni2016, Boeloeni2021}.

  \subsection{Definition of an orographic source}
  \label{subsection_definition_of_an_orographic_source}

    The orographic source is formulated as a lower-boundary condition. For this purpose, the WKB ansatz is applied to the surface topography, i.e.
    \begin{align}
      h \lef(x, y\rig) = h_\mathrm{m} + \sum_\alpha \Re \lef\{h_{\mathrm{w}, \alpha} \exp \lef[i \varphi_\alpha \lef(x, y\rig)\rig]\rig\},
    \end{align}
    where $h_\mathrm{m}$ and $h_{\mathrm{w}, \alpha}$ vary slowly in $x$ and $y$, and $\varphi_\alpha$ are the phases of the small-scale orography. At $z = h$, the wind component orthogonal to the topography must vanish, so that the lower-boundary condition for the flow is given by
    \begin{align}
      0 = \vec{u} \cdot \vec{n} = - u \pdx{h} - \upsilon \pdy{h} + w \qquad \mathrm{at} \quad z = h,
    \end{align}
    where $\vec{n}$ is a corresponding normal vector. Inserting the WKB decomposition of wind and orography yields
    \begin{align}
      \label{lower_boundary_condition_vertical}
      w_{\mathrm{w}, \alpha} = i \vec{k}_\alpha \cdot \vec{u}_\mathrm{m} h_{\mathrm{w}, \alpha} = - i \hat{\omega}_\alpha h_{\mathrm{w}, \alpha} \qquad \mathrm{at} \quad z = h_\mathrm{m},
    \end{align}
    where $w_{\mathrm{w}, \alpha}$ are the gravity-wave amplitudes of the vertical wind. Note that this last step uses the fact that mountain waves are stationary, so that
    \begin{align}
      \label{lower_boundary_condition_frequency}
      \omega_\alpha = \vec{k}_\alpha \cdot \vec{u}_\mathrm{m} + \hat{\omega}_\alpha = 0.
    \end{align}
    As is derived in \citet{Achatz2017, Achatz2023}, the wave-action densities can be written as
    \begin{align}
      \label{wave_action_density}
      \mathcal{A}_\alpha = \frac{\bar{\rho}}{2 \hat{\omega}_\alpha} \lef(\lef|\vec{u}_{\mathrm{w}, \alpha}\rig|^2 + \frac{\lef|b_{\mathrm{w}, \alpha}\rig|^2}{N^2}\rig) = \frac{\bar{\rho}}{2} \frac{\hat{\omega}_\alpha \lef|\vec{k}_\alpha\rig|^2}{N^2 \lef(k_\alpha^2 + l_\alpha^2\rig)} \frac{\lef|b_{\mathrm{w}, \alpha}\rig|^2}{N^2},
    \end{align}
    where $\vec{u}_{\mathrm{w}, \alpha} = \lef(u_{\mathrm{w}, \alpha}, \upsilon_{\mathrm{w}, \alpha}, w_{\mathrm{w}, \alpha}\rig)^\mathrm{T}$ and $b_{\mathrm{w}, \alpha}$ are the gravity-wave amplitudes of the wind and buoyancy, respectively. Using \eqref{phase_space_wave_action_density} and the polarization relation $w_{\mathrm{w}, \alpha} = i \hat{\omega}_\alpha b_{\mathrm{w}, \alpha} / N^2$, one finds that \eqref{lower_boundary_condition_vertical} implies
    \begin{align}
      \label{lower_boundary_condition_wave_action_density}
      \mathcal{N} = \frac{\bar{\rho}}{2} \sum\limits_\alpha \frac{\hat{\omega}_\alpha \lef|\vec{k}_\alpha\rig|^2}{k_\alpha^2 + l_\alpha^2} \lef|h_{\mathrm{w}, \alpha}\rig|^2 \delta \lef(\vec{k} - \vec{k}_\alpha\rig) \qquad \mathrm{at} \quad z = h_\mathrm{m}.
    \end{align}

    The horizontal wavenumbers at the lower boundary are given by the respective derivatives of the orographic phases $\varphi_\alpha$ and the intrinsic frequencies are determined from the latter using \eqref{lower_boundary_condition_frequency}. Lastly, the vertical wavenumbers can be computed from the gravity-wave dispersion relation. In particular, one has
    \begin{align}
      m_\alpha^2 = \frac{\lef(k_\alpha^2 + l_\alpha^2\rig) \lef(N^2 - \hat{\omega}_\alpha^2\rig)}{\hat{\omega}_\alpha^2}.
    \end{align}
    The correct signs are inferred from the fact that the vertical group velocities should be positive at the lower boundary. Since the latter are proportional to $- m_\alpha / \hat{\omega}_\alpha$, the vertical wavenumbers must be
    \begin{align}
      m_\alpha = - \sgn \lef(\hat{\omega}_\alpha\rig) \sqrt{\frac{\lef(k_\alpha^2 + l_\alpha^2\rig) \lef(N^2 - \hat{\omega}_\alpha^2\rig)}{\hat{\omega}_\alpha^2}}.
    \end{align}

  \subsection{Implementation of a Rayleigh damping}
  \label{subsection_implementation_of_a_rayleigh_damping}

    With an orographic source at the lower boundary, wave energy will be transported upwards and eventually reach the upper boundary of the considered domain. In the wave-resolving simulations conducted in the present study, a Rayleigh damping is used to prevent reflections at that boundary. This damping is given by
    \begin{align}
      \label{rayleigh_damping}
      \lef(\pdt{\vec{u}}\rig)_\mathrm{R} & = - \alpha_\mathrm{R} \lef(\vec{u} - \bar{\vec{u}}\rig), & \lef(\pdt{b}\rig)_\mathrm{R} & = - \alpha_\mathrm{R} b,
    \end{align}
    where $\alpha_\mathrm{R} \lef(z\rig)$ is the height-dependent damping coefficient, $\bar{\vec{u}}$ denotes the horizontally averaged wind and $b$ is the buoyancy. In the WKB simulations, the wave absorption is mimicked with a corresponding sink on the right-hand side of \eqref{phase_space_wave_action_conservation}. Using \eqref{wave_action_density} and \eqref{phase_space_wave_action_density}, one finds that the appropriate form of this sink is
    \begin{align}
      \mathcal{S}_1 = - 2 \alpha_\mathrm{R} \mathcal{N}.
    \end{align}

    \subsection{Steady-state theory}
    \label{subsection_steady_state_theory}

      As most operational gravity-wave parameterizations are based on a single-column steady-state WKB theory \citep{Fritts2003}, it is sensible to have a corresponding reference for the model evaluation. Reducing the present theory accordingly has the following implications. In the phase-space-wave-action-density equation \eqref{phase_space_wave_action_conservation}, only derivatives with respect to $z$ and $m$ remain, so that integrating over spectral space yields
      \begin{align}
        \label{phase_space_wave_action_conservation_steady_state}
        \pdz{} \int c_{\mathrm{g} z} \mathcal{N} \, \ddd V_\vec{k} = \int \sum\limits_s \mathcal{S}_s \, \ddd V_\vec{k},
      \end{align}
      where it has been assumed that $\mathcal{N} \rightarrow 0$ for $\lef|m\rig| \rightarrow \infty$. Moreover, \eqref{phase_space_wave_action_density} is valid in a stationary sense, i.e. the number of spectral modes that describe the full wave field remains constant. Because of the quasilinear approximation that is implied in \eqref{phase_space_wave_action_conservation}, each of these modes satisfies
      \begin{align}
        \label{wave_action_conservation_steady_state}
        \pdz{} \lef(c_{\mathrm{g} z, \alpha} \mathcal{A}_\alpha\rig) = \sum\limits_s \mathcal{Q}_{s, \alpha},
      \end{align}
      where $\mathcal{Q}_{s, \alpha}$ are the corresponding sources and sinks, such that $\mathcal{S}_s = \sum_\alpha \mathcal{Q}_{s, \alpha} \delta \lef(\vec{k} - \vec{k}_\alpha\rig)$. Note that the dissipation terms $\mathcal{Q}_{0, \alpha}$ are computed with a turbulent viscosity and diffusivity that takes the full spectrum into account (see below).

      Solving \eqref{wave_action_conservation_steady_state} requires information about the local wavenumbers and frequencies, which are described by the eikonal equations \eqref{ray_equations_frequency}-\eqref{ray_equations_wavenumbers}. These become
      \begin{align}
        c_{\mathrm{g} z, \alpha} \pdz{\omega_\alpha} & = 0,\\
        c_{\mathrm{g} z, \alpha} \pdz{\vec{k}_\alpha} & = \lef(0, 0, - \pdz{\Omega_\alpha}\rig)^\mathrm{T},
      \end{align}
      which means that the extrinsic frequencies and horizontal wavenumbers at all levels are prescribed by the source. The local intrinsic frequencies and vertical wavenumbers can therefore be inferred from the latter, following
      \begin{align}
        \label{steady_state_intrinsic_frequency}
        \hat{\omega}_\alpha & = - \vec{k}_\alpha \cdot \vec{u}_\mathrm{m},\\
        \label{steady_state_vertical_wavenumber}
        m_\alpha & = - \sgn \lef(\hat{\omega}_\alpha\rig) \sqrt{\frac{\lef(k_\alpha^2 + l_\alpha^2\rig) \lef(N^2 - \hat{\omega}_\alpha^2 \rig)}{\hat{\omega}_\alpha^2}}.
      \end{align}

      To ensure that the dissipation of wave-action density due to saturation at any given level is consistent with the corresponding temporal change taking place in the transient model, $\mathcal{Q}_{0, \alpha}$ is integrated over the pseudo-time step $\Delta z / c_{\mathrm{g} z, \alpha}$. Consequently, the turbulent viscosity and diffusivity of the steady-state model is given by
      \begin{align}
        \label{diffusion_coefficient_steady_state}
        D = \frac{\bar{\rho}}{4} \lef[\sum\limits_\alpha \frac{\Delta z}{c_{\mathrm{g} z, \alpha}} N^4 \lef(k_\alpha^2 + l_\alpha^2\rig) m_\alpha^2 \frac{\mathcal{A}_\alpha}{\hat{\omega}_\alpha}\rig]^{- 1} \max \lef[0, \frac{2}{\bar{\rho}} \sum\limits_\alpha \frac{N^4 \lef(k_\alpha^2 + l_\alpha^2\rig) m_\alpha^2}{\hat{\omega}_\alpha \lef|\vec{k}_\alpha\rig|^2} \mathcal{A}_\alpha - \alpha_\mathrm{d}^2 N^4\rig].
      \end{align}
      Note that this has the effect of waves with small group velocities dissipating more strongly \citep{Boeloeni2021}.

      Because the horizontal gravity-wave momentum fluxes are assumed to be zero, the mean-flow impact vanishes in the absence of dissipation and Rayleigh damping. However, if the right-hand side of \eqref{wave_action_conservation_steady_state} is nonzero, the mean flow experiences a drag given by
      \begin{align}
        \label{mean_flow_impact_steady_state}
        \lef(\pdt{u_\mathrm{m}}, \pdt{\upsilon_\mathrm{m}}\rig)_\mathrm{w} & = - \frac{1}{\bar{\rho}} \pdz{} \sum\limits_\alpha c_{\mathrm{g} z, \alpha} \lef(k_\alpha, l_\alpha\rig) \mathcal{A}_\alpha
      \end{align}
      \citep{McFarlane1987, Boeloeni2016, Boeloeni2021}.

      A special case that needs to be considered is the occurrence of critical or reflecting levels. At a critical level, the intrinsic frequency approaches zero (in the absence of rotation), causing the vertical wavenumber to diverge. Consequently, the vertical propagation slows down and the waves are ultimately trapped beneath the level. This means that the local wave-action density must be set to zero above the critical level. At a reflecting level, the intrinsic frequency approaches the buoyancy frequency, resulting in the vertical wavenumber becoming smaller in magnitude and finally flipping its sign. Thus, waves cannot propagate beyond such a level either (being reflected instead), implying zero wave action above it \citep{Boeloeni2021}.

\section{Results}
\label{section_results}

  In this section, the results from several idealized simulations conducted with the described model are discussed and compared to reference data from wave-resolving simulations with PincFlow.

  \subsection{Configuration}
  \label{subsection_configuration}

    PincFlow \citep{Rieper2013, Schmid2021} integrates the pseudo-incompressible equations in a conservative flux form \citep{Klein2009}, using a semi-implicit time scheme. Its spatial discretization follows a finite-volume method with a monotone upwind scheme \citep{Leer2003} for mass and momentum fluxes. The flow solver uses an adaptive time step computed from several stability criteria, including CFL constraints with respect to the resolved flow and the group velocities determined by the WKB model. PincFlow's coordinate system is terrain-following \citep[based on][]{GalChen1975}, defined from either the full orography (in wave-resolving simulations) or the orographic background (in WKB simulations). In the simulations run for the present study, the prognostic equations of the WKB model (i.e. \eqref{phase_space_wave_action_conservation} and \eqref{phase_space_rays}) are integrated with a third-order Runge-Kutta scheme \citep{Williamson1980} at the beginning of each time step. The corresponding mean-flow impact (i.e. \eqref{mean_flow_impact}) is then added to the momentum equation and treated as a constant forcing for the rest of the time step.

    \begin{table}
      \caption[Simulation parameters]{Simulation parameters. The grid parameters of the WKB simulations are written in parentheses.}
      \label{table_simulation_parameters}
      \begin{center}
        \begin{tabular}{lll}
          \topline
          Parameter & Value & Description\\
          \midline
          $L_x$ & $60 \, \mathrm{km}$ & Horizontal extent of the domain\\
          $L_z$ & $100 \, \mathrm{km}$ & Vertical extent of the domain\\
          $N_x$ & $192$ ($3$) & Grid points in horizontal direction\\
          $N_z$ & $1920$ ($240$) & Grid points in vertical direction\\
          $N_0$ & $0.0179 \, \mathrm{s^{- 1}}$ & Buoyancy frequency\\
          $u_0$ & $10 \, \mathrm{m \, s^{- 1}}$ & Initial horizontal wind\\
          $z_\mathrm{R}$ & $9 \, \mathrm{km}$ & Rayleigh-damping scale height\\
          $\alpha_{\mathrm{R}, \max}$ & $0.0179 \, \mathrm{s^{- 1}}$ & Maximum Rayleigh damping\\
          $h_0$ & $100$-$1000 \, \mathrm{m}$ & Mountain heights\\
          $l_0$ & $10 \, \mathrm{km}$ & Mountain half width\\
          $\alpha_\mathrm{d}$ & $1.0$ & Wave-saturation parameter\\
          \botline
        \end{tabular}
      \end{center}
    \end{table}

    The conducted simulations cover flows ranging from the linear to the nonlinear regime, in an idealized 2D setting. This setting features an isothermal atmosphere and a monochromatic orography given by
    \begin{align}
      h \lef(x\rig) = \frac{h_0}{2} \lef[1 + \cos \lef(\frac{\pi x}{l_0}\rig)\rig],
    \end{align}
    such that $h_\mathrm{m} = h_\mathrm{w} = h_0 / 2$ and $k_h = \pi / l_0$. The domain is horizontally periodic and has a rigid upper boundary. To prevent wave reflections at the latter, the wind and buoyancy fluctuations are damped following \eqref{rayleigh_damping} (where $\bar{\vec{u}}$ is the mean along the terrain-following coordinate lines), with the Rayleigh-damping coefficient
    \begin{align}
      \alpha_\mathrm{R} \lef(z\rig) = \alpha_{\mathrm{R}, \max} \exp \lef(\frac{z - L_z}{z_\mathrm{R}}\rig)
    \end{align}
    \citep{Prusa2003}. All simulations have an integration time of one day, over the first three hours of which the orography is grown linearly from zero to its maximum height. This method ensures a gentle initialization in the sense that the orographic source is always linear at first. In the WKB simulations, this is done by growing both the wave amplitude $h_\mathrm{w}$ and the orographic background $h_\mathrm{m}$. Note that replacing the orography growth with an initial acceleration of the wind would result in a nonlinear flow at the beginning of each simulation. On the other hand, a forced deceleration that induces the transition from linear to nonlinear regime would require an initial wind strong enough to generate an evanescent wave over the highest mountains.

    Apart from molecular friction, no explicit diffusion is used. Thus, artificial diffusion is limited to what is intrinsic to the discretization schemes, and turbulence is not being parameterized explicitly. Note that the numerical diffusion of the monotone upwind scheme reduces grid noise. A summary of the relevant simulation parameters is given in Table \ref{table_simulation_parameters}. The individual configurations will be denoted with the following abbreviations.
    \begin{itemize}
      \item WRS: Wave-resolving simulations in a high-resolution setup of the flow solver
      \item WKB-TR: Low-resolution simulations in which the waves are parameterized with the transient WKB model
      \item WKB-ST: Low-resolution simulations in which the waves are parameterized with the steady-state WKB model
    \end{itemize}

  \subsection{Flow decomposition}
  \label{subsection_flow_decomposition}

    To compare the wave-resolving dynamics to the output of the WKB model, the WRS flow must be decomposed. Due to the monochromatic orography, one might expect that a horizontal average should suffice. However, as soon as the flow becomes nonlinear, small-scale wave-like structures appear in the vertical profiles of the horizontally averaged wind (see below). To facilitate the visual comparison of such profiles, a local average defined by
    \begin{align}
      \label{local_average}
      \lef\langle{}a\rig\rangle \lef(\hat{z}, t\rig) & = \frac{N_0}{2 \pi u_0 L_x} \, \int\limits_{- \frac{L_x}{2}}^{\frac{L_x}{2}} \int\limits_{\hat{z} - \frac{\pi u_0}{N_0}}^{\hat{z} + \frac{\pi u_0}{N_0}} a \lef(\hat{x}, \hat{z}, t\rig) \, \ddd \hat{x} \ddd \hat{z}
    \end{align}
    is applied instead, where $a$ is an arbitrary quantity of the resolved flow and $\lef(\hat{x}, \hat{z}\rig)$ are the terrain-following coordinates of the flow solver. Near the vertical boundaries, where $\hat{z} \leq \Delta \hat{z} / 2 +  \pi u_0 / N_0$ or $\hat{z} \geq L_z - \Delta \hat{z} / 2 - \pi u_0 / N_0$, the integrand in \eqref{local_average} is replaced with
    \begin{align}
      \label{mirror}
      \til{a} \lef(\hat{x}, \hat{z}, t\rig) =
      \begin{cases}
        2 a \lef(\hat{x}, \frac{\Delta \hat{z}}{2}, t\rig) - a \lef(\hat{x}, \Delta \hat{z} - \hat{z}, t\rig) & \mathrm{if} \quad \hat{z} < \frac{\Delta \hat{z}}{2},\\
        a \lef(\hat{x}, \hat{z}, t\rig) & \mathrm{if} \quad \frac{\Delta \hat{z}}{2} \leq \hat{z} \leq L_z - \frac{\Delta \hat{z}}{2},\\
        2 a \lef(\hat{x}, L_z - \frac{\Delta \hat{z}}{2}, t\rig) - a \lef(\hat{x}, 2 L_z - \Delta\hat{z} - \hat{z}, t\rig) & \mathrm{if} \quad \hat{z} > L_z - \frac{\Delta \hat{z}}{2}.
      \end{cases}
    \end{align}
    This mirroring technique ensures that the local average converges to a mean along the terrain-following coordinate lines at $\hat{z} = \Delta \hat{z} / 2$ and $\hat{z} = L_z - \Delta \hat{z} / 2$, while maintaining some vertical smoothing.

  \subsection{Blocking}
  \label{subsection_blocking}

    For a proper understanding of the dynamics that are to be emulated by the WKB model, it is sensible to first examine the wave-resolving simulations. With the grid spacings $\Delta \hat{x} = 312.5 \, \mathrm{m}$ and $\Delta \hat{z} \approx 52 \, \mathrm{m}$, the flow solver resolves the orographic wavelength ($2 l_0 = 20 \, \mathrm{km}$) and initial hydrostatic vertical wavelength ($2 \pi u_0 / N_0 \approx 3.5 \, \mathrm{km}$) with approximately $64$ grid points each. Such a high resolution is necessary not only to capture wave breaking but also to resolve smaller vertical wavelengths that are generated after the mean flow has been decelerated. Moreover, to detect nonlinear boundary effects near the surface, it is vital to have a sufficient number of vertical layers below the mountain summits. A well-known phenomenon that falls under this latter category is blocking, which occurs when the amplitude of a generated mountain wave is large enough to prevent the mean flow in valleys from crossing the summits. In such a situation, the stagnant air that accumulates in the valleys acts as an effective lower boundary, so that the generated wave amplitude is reduced \citep{Lott1997, Lott1999}. Since the range of mountain heights considered in the present study reaches well into the nonlinear regime, it can be expected that such a flow development is observable in some of the wave-resolving simulations.

    \begin{figure*}
      \includegraphics[width = \textwidth]{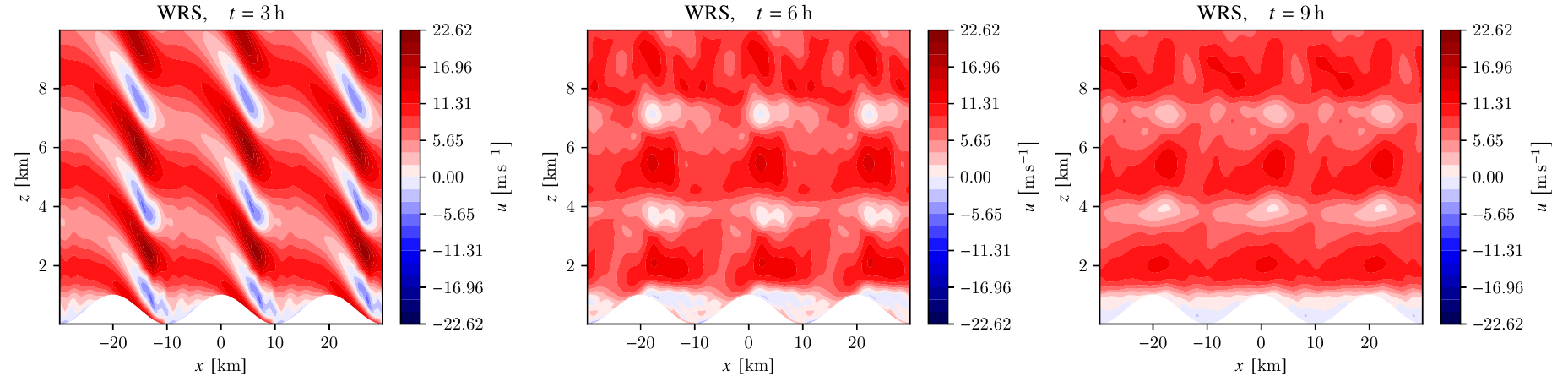}
      \caption{Horizontal wind below $10 \, \mathrm{km}$, as simulated in WRS for $h_0 = 1000 \, \mathrm{m}$. The shown snapshots are $3 \, \mathrm{h}$ (left), $6 \, \mathrm{h}$ (center) and $9 \, \mathrm{h}$ (right) after initialization.}
      \label{figure_horizontal_wind}
    \end{figure*}

    To illustrate this, Fig. \ref{figure_horizontal_wind} shows snapshots of the horizontal wind simulated below $10 \, \mathrm{km}$ for $h_0 = 1000 \, \mathrm{m}$. After $3 \, \mathrm{h}$, the orography has fully grown and the generated mountain wave is becoming unstable. As it breaks, it decelerates the flow at $\sim 4 \, \mathrm{km}$ and $\sim 7 \, \mathrm{km}$, as can be seen $3 \, \mathrm{h}$ later. Furthermore, a strong deceleration in the valleys sets in, due to a combination of wave-mean-flow interaction and blocking. After $9 \, \mathrm{h}$, the flow is slowly becoming steady, with an almost purely vertical dependence and no more wave generation at the lower boundary.

    Note that the impact of blocking has not been considered in the above description of the WKB model. Many operational mountain-wave parameterizations \citep[see][]{Niekerk2020} represent this phenomenon with variations of the scheme developed by \citet{Lott1997} and \citet{Lott1999}. This scheme emulates the mean-flow deceleration in valleys with an approximative drag that is determined from three-dimensional orographic geometry and an estimation for the depth of the blocked layer, dependent on a non-dimensional mountain height. Moreover, it computes the mountain-wave drag from an effective orography, the minimum of which is given by the upper edge of the blocked layer. However, the blocking in WRS is different from the situation considered by classic parameterizations, as there is no distinction between upstream and downstream flow, due to the horizontal periodicity. The mean flow near the surface simultaneously experiences blocking and deceleration due to interaction with gravity waves at every point along the horizontal axis. From Fig. \ref{figure_horizontal_wind} alone, it is unclear how much each of these two effects contributes to the overall deceleration of the near-surface flow, although it is to be expected that wave-mean-flow interaction plays a more important role than it would for an isolated mountain range. To focus on this aspect of the dynamics and avoid the potential uncertainties of a blocked-layer scheme, the orographic source of the WKB model is left unchanged in the present study.

  \subsection{Low mountains}
  \label{subsection_low_mountains}

    \begin{figure*}[b]
      \includegraphics[width = \textwidth]{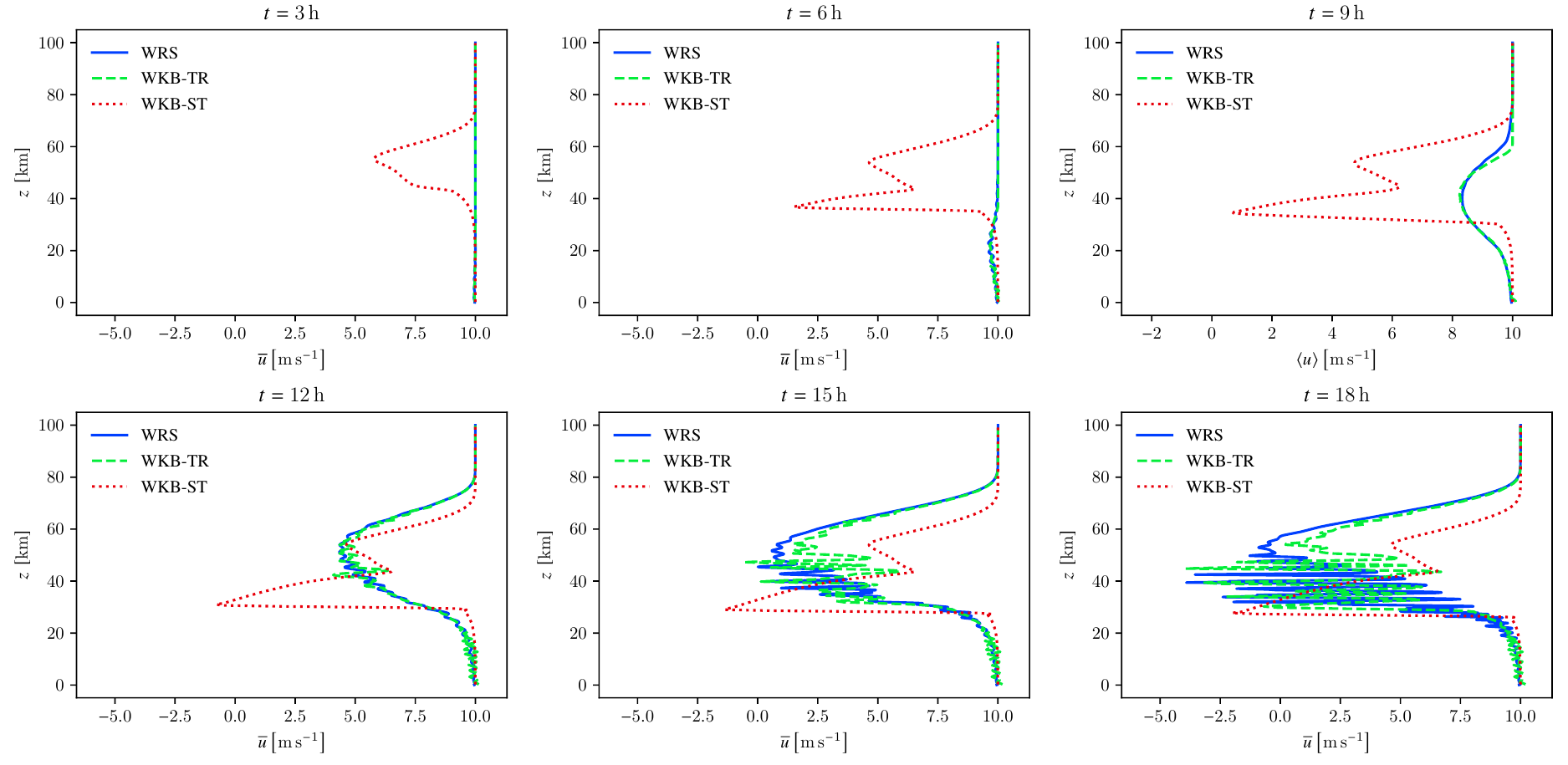}
      \caption{Mean-wind profiles from the WRS (blue), WKB-TR (green) and WKB-ST (red) simulations, with $h_0 = 100 \, \mathrm{m}$. The shown snapshots are $3 \, \mathrm{h}$ (upper left), $6 \, \mathrm{h}$ (upper center), $9 \, \mathrm{h}$ (upper right), $12 \, \mathrm{h}$ (lower left), $15 \, \mathrm{h}$ (lower center) and $18 \, \mathrm{h}$ (lower right) after initialization. The mean wind was computed by averaging along the terrain-following coordinate lines.}
      \label{figure_mean_wind_profiles_linear_horizontal_mean}
    \end{figure*}

    Before nonlinear mountain-wave generation is considered, it is important to verify the model's capability of reproducing a flow that is linear at low altitudes. Fig. \ref{figure_mean_wind_profiles_linear_horizontal_mean} displays snapshots of the mean wind simulated in WRS, WKB-TR and WKB-ST for $h_0 = 100 \, \mathrm{m}$. The profiles were computed by averaging along the terrain-following coordinate lines. It is to be noted that a truly horizontal average (i.e. in Cartesian coordinates) produces nearly identical results (not shown). After $3 \, \mathrm{h}$, the orography has fully grown, so that the emitted wave energy from this point onward remains relatively constant (since the mean wind does not change significantly at the surface). The mean winds simulated in WRS and WKB-TR are still almost unchanged, whereas that in WKB-ST has been decelerated substantially between $\sim 30 \, \mathrm{km}$ and $\sim 70 \, \mathrm{km}$. This difference visualizes the error in the assumption of instantaneous wave-energy distribution. In WKB-ST, the wave signal of a stratified flow crossing mountains with $h_0 = 100 \, \mathrm{m}$ has already affected the entire domain. In fact, the steady-state model already predicts static instability. Because the mean wind is mainly affected by the momentum-flux convergence due to the Rayleigh damping, the impact of wave breaking is only barely noticeable from small irregularities in the profile between $\sim 40 \, \mathrm{km}$ and $\sim 60 \, \mathrm{km}$. At higher altitudes, the Rayleigh damping becomes dominant, reducing the wave amplitude sufficiently to counteract the anelastic amplification and prevent wave breaking. In contrast to WKB-ST, the wave is still propagating upwards in WRS. The transient model can capture this slow development, thus predicting a more realistic mean wind at high altitudes.

    At $6 \, \mathrm{h}$, the wave saturation in WKB-ST has led to a local maximum of $\bar{u}$ between $\sim 40 \, \mathrm{km}$ and $\sim 50 \, \mathrm{km}$. Meanwhile, the mean wind in WRS has been decelerated below $\sim 40 \, \mathrm{km}$. The transient model produces a similar forcing, resulting in a matching profile. Notably, the steady-state model underestimates the forcing below $\sim 30 \, \mathrm{km}$. This is a clear indication of transience being dominant at those altitudes. In the following hours, the mean wind in WRS is decelerated further as the wave propagates upwards. A similar development can be seen in WKB-TR, whereas the mean wind in WKB-ST approaches a steady state.

    \begin{figure*}
      \includegraphics[width = \textwidth]{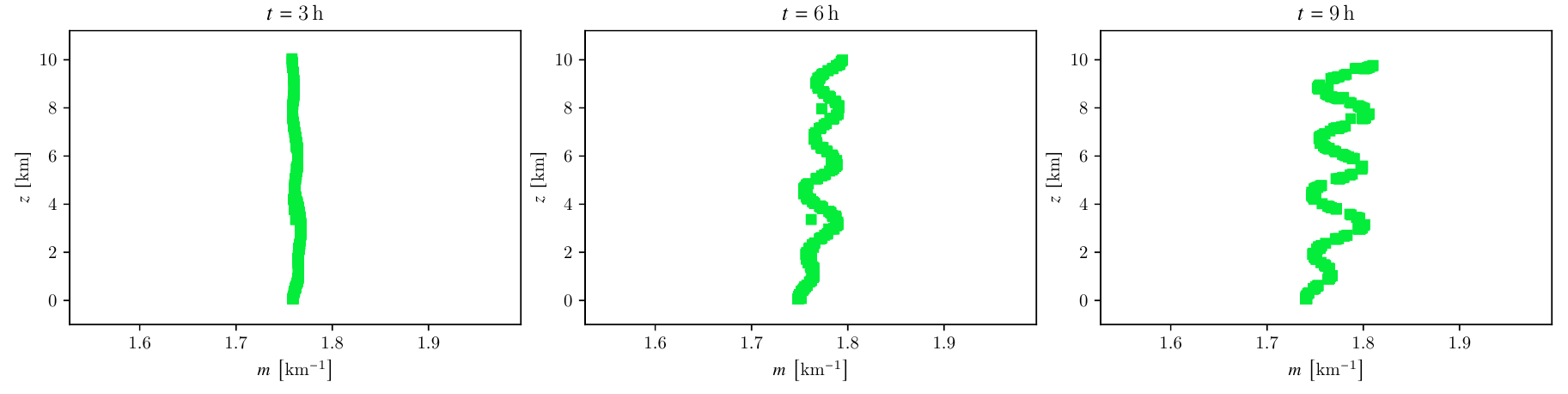}
      \caption{Distribution of the ten dominant ray volumes in each grid cell below $10 \, \mathrm{km}$, as simulated in WKB-TR for $h_0 = 100 \, \mathrm{m}$. The shown snapshots are $3 \, \mathrm{h}$ (left), $6 \, \mathrm{h}$ (center) and $9 \, \mathrm{h}$ (right) after initialization.}
      \label{figure_ray_volumes}
    \end{figure*}

    Not until $15 \, \mathrm{h}$ after initialization does wave breaking become noticeable in the WRS and WKB-TR mean winds (although it sets in earlier). The transient model approximates the corresponding reduction of wave energy with the saturation scheme described in section \ref{section_theory}\ref{subsection_wave_breaking}. The uncertainties therein (represented by the parameter $\alpha_\mathrm{d}$) lead to the first substantial differences between WRS and WKB-TR. However, the transient model remains notably closer to the reference, especially below $\sim 30 \, \mathrm{km}$ and above $\sim 60 \, \mathrm{km}$, where no wave saturation has occurred. Most notably, the wave breaking leads to small-scale oscillations in the mean wind, with a wavelength close to $2 \pi u_0 / N_0$. Although WKB theory does not directly describe these structures, the transient model is able to replicate them in terms of amplitude and location. On the other hand, the mean wind forced by the steady-state model becomes relatively stationary after $12 \, \mathrm{h}$. This can be explained with the fact that a critical level forms around this time, as the mean wind becomes negative at $\sim 30 \, \mathrm{km}$. Below this critical level, the saturation criterion is not fulfilled and the Rayleigh damping is small, so that the mean wind remains virtually constant. Above it, the wave action vanishes by the definition of critical levels and steady-state theory. Note that this does not happen in WKB-TR, since critical levels vanish in transient WKB theory \citep[see][]{Senf2011, Broutman1986}. As can be seen in the snapshot at $18 \, \mathrm{h}$, the mean wind even becomes negative at multiple levels.

    \begin{figure*}
      \includegraphics[width = \textwidth]{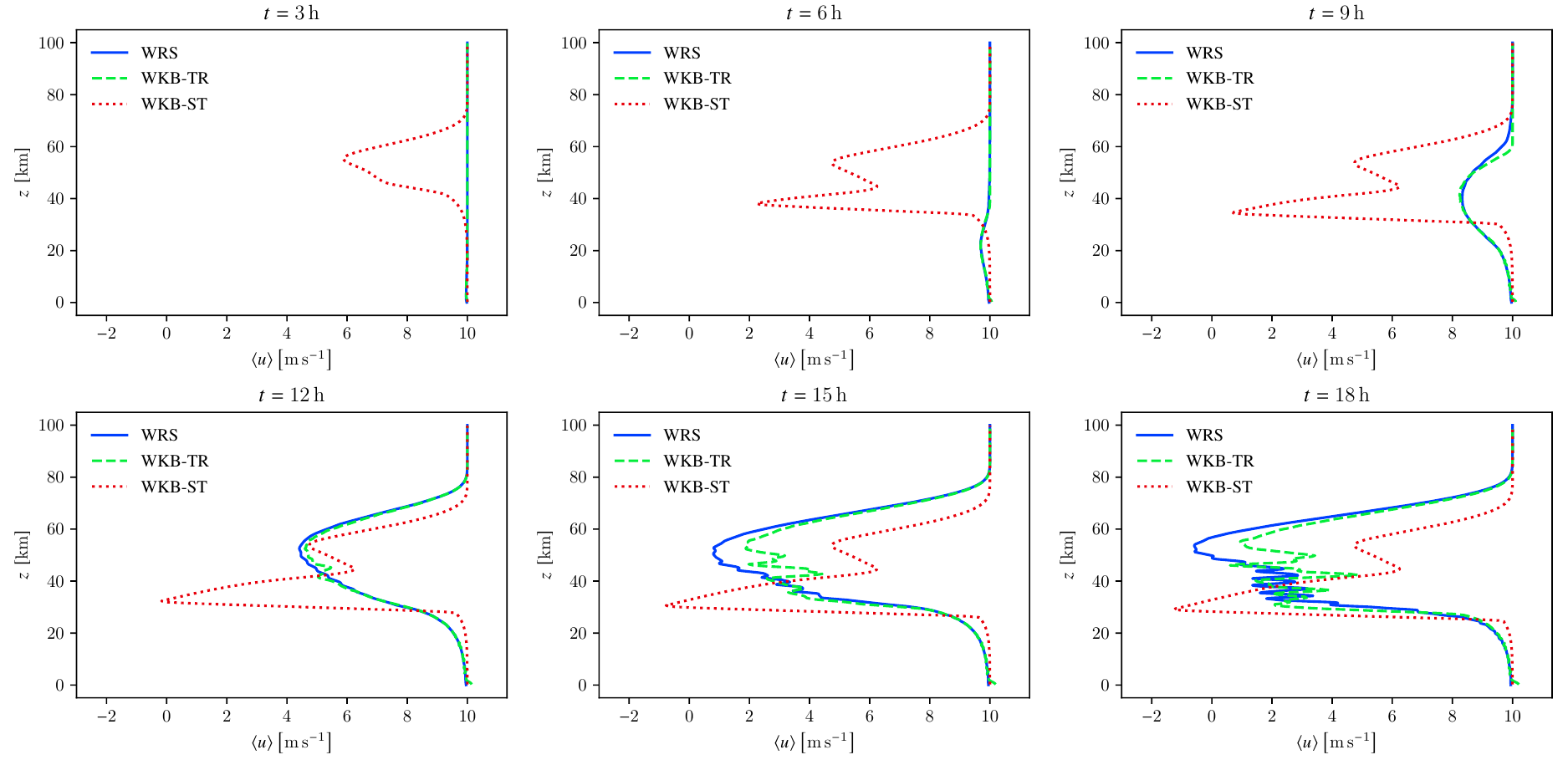}
      \caption{Mean-wind profiles from the WRS (blue), WKB-TR (green) and WKB-ST (red) simulations, with $h_0 = 100 \, \mathrm{m}$. The shown snapshots are $3 \, \mathrm{h}$ (upper left), $6 \, \mathrm{h}$ (upper center), $9 \, \mathrm{h}$ (upper right), $12 \, \mathrm{h}$ (lower left), $15 \, \mathrm{h}$ (lower center) and $18 \, \mathrm{h}$ (lower right) after initialization. The mean wind was computed using \eqref{local_average}.}
      \label{figure_mean_wind_profiles_linear}
    \end{figure*}

    The mean-wind oscillations in WRS and WKB-TR are reminiscent of modulational instabilities. These can arise from perturbations of a saturated, stationary solution to the locally monochromatic limit of \eqref{ray_equations_wavenumbers}, \eqref{phase_space_wave_action_conservation} and \eqref{mean_flow_impact} \citep[see e.g.][]{Schlutow2019}. In WRS and WKB-TR, the conditions for such a situation are given. The wave field is locally monochromatic and eventually saturates above a certain level (at which point the spectrum broadens). Moreover, the Rayleigh damping represents an additional wave-action sink below saturation. In fact, mean-wind oscillations even occur at levels where the wave is stable, as can be seen in the snapshot at $12 \, \mathrm{h}$. Further insight can be gained by looking at the evolution of the vertical wavenumber in WKB-TR. Fig. \ref{figure_ray_volumes} shows snapshots of the ten most dominant ray volumes of each grid cell in the $z$-$m$ plane below $10 \, \mathrm{km}$. One can clearly see how they initially propagate along a straight path, which becomes wave-like as the mountain wave interacts with the mean flow. However, this modulation appears to be stationary, which suggests that it is not an upwards-propagating perturbation. It might be a related phenomenon intrinsic to the above-mentioned equation system.

    Although the wave-like structures in the WRS and WKB-TR mean winds are a noteworthy observation, they also make it difficult to distinguish the vertical profiles. For this reason, the horizontal average will be replaced by the local average \eqref{local_average} in the following, both to compute the mean-flow quantities of WRS and to smooth the output of the WKB simulations. Fig. \ref{figure_mean_wind_profiles_linear} illustrates how this affects the profiles for $h_0 = 100 \, \mathrm{m}$. Note that some mean-wind oscillations are still visible in the snapshots at $15 \, \mathrm{h}$ and $18 \, \mathrm{h}$ but their amplitudes are reduced.

    \begin{figure*}
      \includegraphics[width = \textwidth]{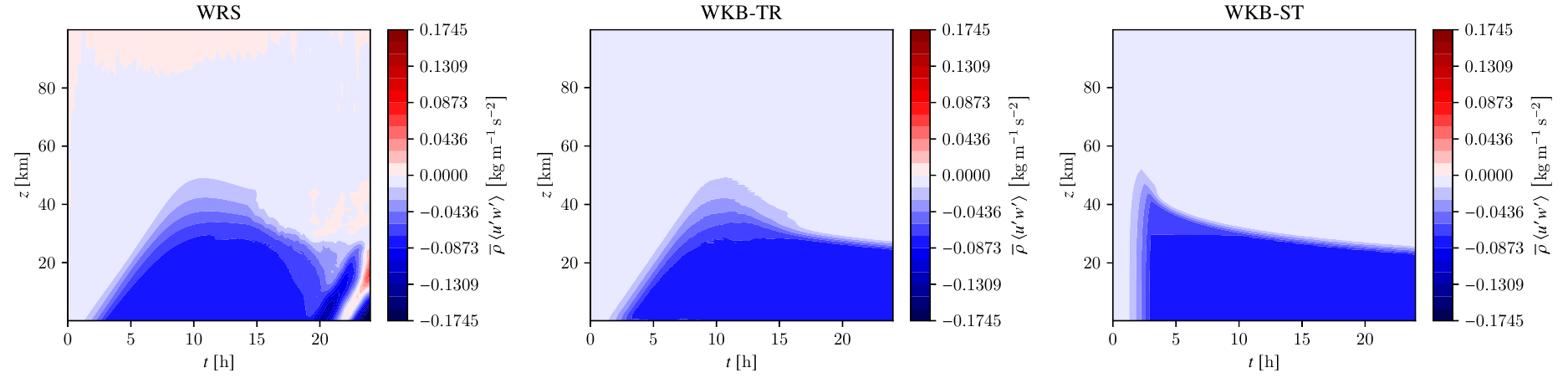}
      \caption{Hovmoeller plots of the momentum flux in WRS (left), WKB-TR (center) and WKB-ST (right), for $h_0 = 100 \, \mathrm{m}$. The momentum flux was computed using \eqref{local_average}.}
      \label{figure_momentum_flux}
    \end{figure*}

    One way to gain a better idea of how and when wave breaking affects the flow is to look at the momentum flux. Fig. \ref{figure_momentum_flux} shows Hovmoeller plots of $\bar{\rho} \lef\langle u' w'\rig\rangle$, for $h_0 = 100 \, \mathrm{m}$. In the wave-resolving simulation, the momentum flux begins to decrease approximately $12 \, \mathrm{h}$ after initialization, indicating saturation. In WKB-TR, this happens slightly earlier, at $\sim 11 \, \mathrm{h}$. Starting at $\sim 15 \, \mathrm{h}$, the differences between WRS and WKB-TR become substantial. The transient WKB model predicts a slow decrease in the momentum flux, whereas the reference shows a rather quick reduction. After $\sim 22 \, \mathrm{h}$, the WRS momentum flux even becomes positive, suggesting the generation of secondary gravity waves. However, the momentum flux in WKB-ST already reaches its peak after $3 \, \mathrm{h}$ and then decreases rapidly above $\sim 30 \, \mathrm{km}$. This is a clear consequence of the instantaneous distribution of wave energy. Overall, the differences in the momentum flux are consistent with those in the mean wind and thus highlight the fact that the interaction between mountain waves and mean flow is a highly transient process that is severely simplified in classic parameterizations.

  \subsection{High mountains}
  \label{subsection_high_mountains}

    \begin{figure*}[b]
      \includegraphics[width = \textwidth]{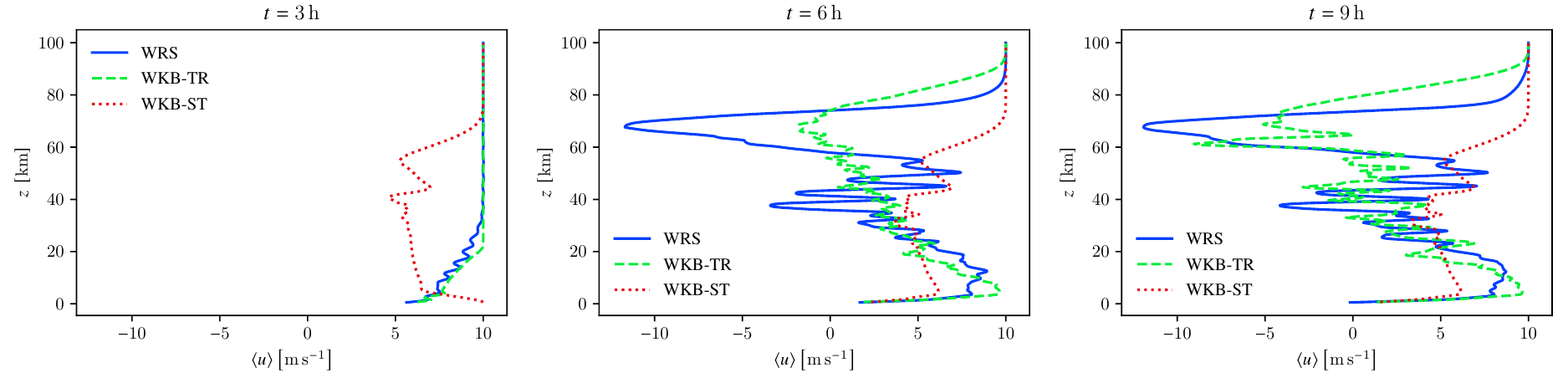}
      \caption{Mean-wind profiles from the WRS (blue), WKB-TR (green) and WKB-ST (red) simulations, with $h_0 = 1000 \, \mathrm{m}$. The shown snapshots are $3 \, \mathrm{h}$ (left), $6 \, \mathrm{h}$ (center) and $9 \, \mathrm{h}$ (right) after initialization. The mean wind was computed using \eqref{local_average}.}
      \label{figure_mean_wind_profiles_nonlinear}
    \end{figure*}

    Having discussed the interaction between the mean flow and a linearly generated mountain wave, a situation where the wave already becomes nonlinear at the lower boundary is to be considered next. Fig. \ref{figure_mean_wind_profiles_nonlinear} shows snapshots of the mean wind simulated for $h_0 = 1000 \, \mathrm{m}$. At $3 \, \mathrm{h}$, the generated mountain wave is becoming unstable near the surface, as can be seen in the left panel of Fig. \ref{figure_horizontal_wind}. As a result of the large wave amplitude, the mean winds in WRS and WKB-TR are decelerated at low altitudes but have not yet been impacted at high altitudes. In contrast, the instantaneous wave-energy distribution in WKB-ST has led to a forcing at all altitudes below $\sim 70 \, \mathrm{km}$. Close to the surface, the forcing of the steady-state model is weaker than that in WRS and WKB-TR, indicating significant transient interaction.

    At $6 \, \mathrm{h}$, the highly nonlinear wave signal in WRS has propagated to the top of the domain, exerting a forcing on the mean wind that is significantly larger than in WKB-ST. The transient model produces a similar tendency, resulting in a mean-wind profile that is relatively close to the reference. Especially noteworthy is the fact that the deceleration near the surface is reproduced in WKB-TR and WKB-ST, although no blocked-layer parameterization is being used. This indicates that the near-surface flow is dominated by wave-mean-flow interaction, with blocking having a comparatively small impact, due to the horizontal periodicity. It is to be pointed out that more realistic simulations (with distinct upstream and downstream flows) will likely exhibit different dynamics. Note also that between $\sim 5 \, \mathrm{km}$ and $\sim 20 \, \mathrm{km}$, the mean wind in WKB-ST is too weak, whereas WKB-TR appears to be comparatively realistic.

    At $9 \, \mathrm{h}$, the mean wind in WRS has been further decelerated near the surface but changed little at higher altitudes. Meanwhile, the mean wind forced by the steady-state model has remained virtually constant. In WKB-TR, the mean wind above $\sim 40 \, \mathrm{km}$ has become more negative and thus closer to that in WRS. This is particularly interesting because the steady-state model is by definition incapable of forcing such a wind reversal, as has been mentioned above. The occurrence of strong negative winds therefore points to a high impact of transience. In the following hours (not shown), the profiles do not change significantly.

  \subsection{Root-mean-square errors}
  \label{subsection_root_mean_square_errors}

    \begin{figure}
      \includegraphics[width = \columnwidth]{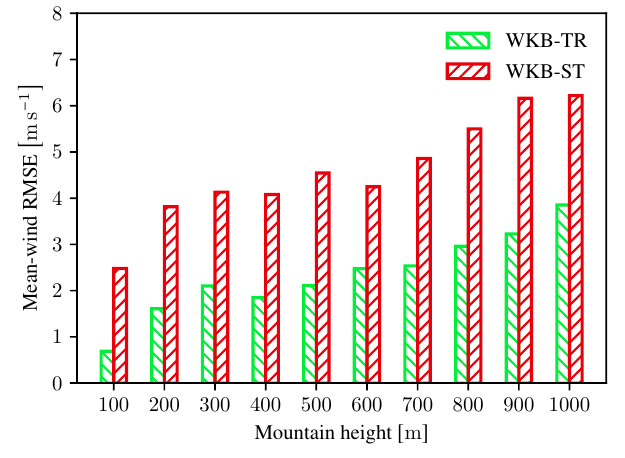}
      \caption{Mean-wind RMSEs in several of the WKB-TR and WKB-ST runs, with respect to the WRS reference (interpolated to the WKB levels). The errors were computed using \eqref{rmse}.}
      \label{figure_rmse_mean_wind}
    \end{figure}

    The large differences between the transient and steady-state versions of the model are especially apparent in the mean-wind root-mean-square errors (RMSEs) with respect to the WRS results. Fig. \ref{figure_rmse_mean_wind} depicts these for several mountain heights. The RMSEs were computed from
    \begin{align}
      \label{rmse}
      \mathrm{RMSE} \lef(\lef\langle u \rig\rangle\rig) = \sqrt{\lef(N_z N_t\rig)^{- 1} \sum\limits_k^{N_z} \sum\limits_n^{N_t} \lef(\lef\langle u \rig\rangle_k^n - \lef\langle u \rig\rangle_{\mathrm{r}, k}^n\rig)^2},
    \end{align}
    where $N_t = 24 \, \mathrm{h} / 15 \, \mathrm{min} + 1$ is the number of outputs and $\lef\langle u \rig\rangle_{\mathrm{r}, k}^n$ are the WRS mean winds interpolated to the WKB levels. In all cases, the WKB-ST RMSEs are larger than the WKB-TR ones. The relative differences are significant, being even greater than $100 \%$ (with respect to the transient model) for $h_0 = 100 \, \mathrm{m}$. Note that the RMSEs are not monotonically increasing with $h_0$. This is likely related to when the low-level flow becomes nonlinear and how quickly it approaches a steady state, which also depends on the orography growth time.

  \subsection{Sensitivity to orography growth time}
  \label{subsection_sensitiviy_to_orography_growth_time}

\begin{figure*}
      \includegraphics[width = \textwidth]{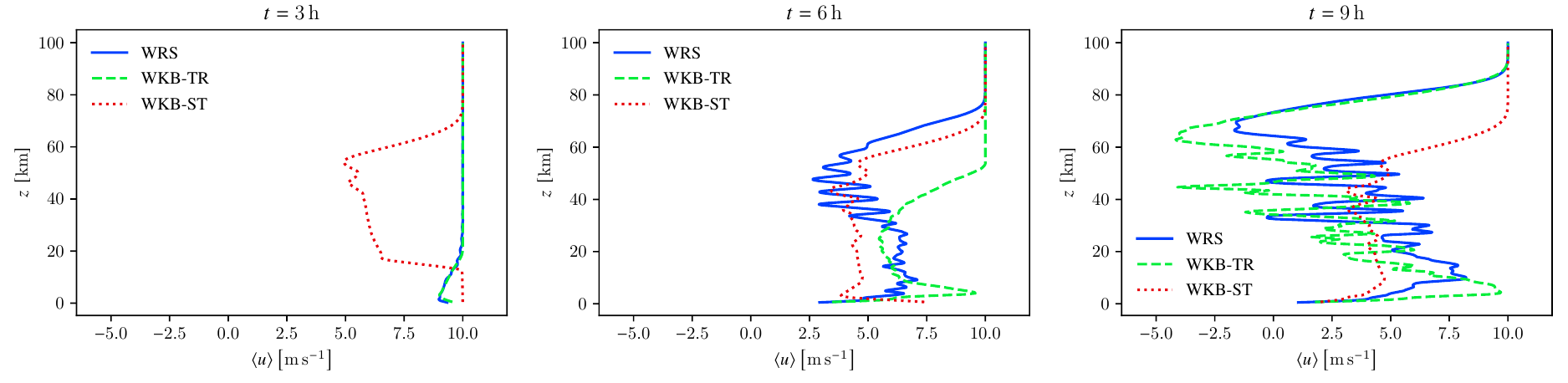}
      \caption{Mean-wind profiles from the WRS (blue), WKB-TR (green) and WKB-ST (red) simulations, with $h_0 = 1000 \, \mathrm{m}$ and an orography growth time of $6 \, \mathrm{h}$. The shown snapshots are $3 \, \mathrm{h}$ (left), $6 \, \mathrm{h}$ (center) and $9 \, \mathrm{h}$ (right) after initialization. The mean wind was computed using \eqref{local_average}.}
      \label{figure_mean_wind_profiles_comparison}
    \end{figure*}

    One might argue that growing the orography over $3 \, \mathrm{h}$ introduces an unphysical transience to the system that complicates the comparison to reality. As has been mentioned in section \ref{section_results}\ref{subsection_configuration}, the main reason for growing the orography is a gentle initialization of the WKB model. Because the horizontal wind is initialized with $u_0 = 10 \, \mathrm{m \, s^{- 1}}$, the orographic source goes through a transition from the linear to the nonlinear regime. This is somewhat comparable to an upstream flow that decelerates. The additional transience is therefore not entirely unrealistic. However, the question arises, how the flow development is impacted by the growth time. The $3 \, \mathrm{h}$ chosen so far correspond to a relatively fast initialization, so that the flow for $h_0 = 1000 \, \mathrm{m}$ becomes nonlinear fairly quickly. A longer growth time leads to a slower, less transient development. Fig. \ref{figure_mean_wind_profiles_comparison} shows the same snapshots as Fig. \ref{figure_mean_wind_profiles_nonlinear} but for an orography growth time of $6 \, \mathrm{h}$. Most notably, the mean winds in WRS and WKB-TR become less negative at high altitudes. Consequently, WKB-ST is closer to the reference than for a growth time of $3 \, \mathrm{h}$. This illustrates how decreasing the growth time results in a less transient flow. In fact, in may be noted here that the additional transience is maximized in the limit of not growing the orography.

    \begin{figure}
      \includegraphics[width = \columnwidth]{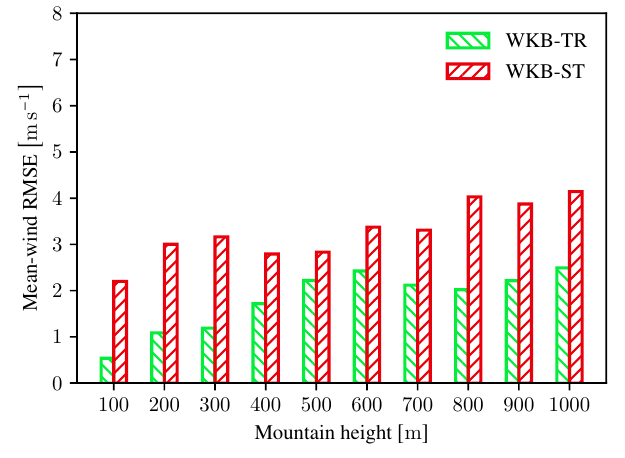}
      \caption{Mean-wind RMSEs in several of the WKB-TR and WKB-ST runs where the orography growth time is $6 \, \mathrm{h}$, with respect to the WRS reference (interpolated to the WKB levels). The errors were computed using \eqref{rmse}.}
      \label{figure_rmse_mean_wind_comparison}
    \end{figure}

    A more complete picture can be gained by looking at the corresponding mean-wind RMSEs. Fig. \ref{figure_rmse_mean_wind_comparison} depicts these, once again computed with \eqref{rmse}. Consistent with the differences between the WKB-ST and WRS mean wind profiles, the WKB-ST RMSEs are smaller than for a growth time of $3 \, \mathrm{h}$. This improvement is largest for high mountains. Notably, WKB-TR is also closer to the reference and once again more realistic than WKB-ST for all $h_0$. This underlines the fact that all the physics included in the steady-state model are also accounted for in the transient model, so that the latter provides a more complete description of the dynamics. Note also that the RMSEs' dependence on $h_0$ is different than in Fig. \ref{figure_rmse_mean_wind}, emphasizing its relation to the growth time. More specifically, growing the orography more slowly delays the time at which the low-level flow becomes nonlinear for intermediate mountain heights. Because the largest differences occur when the wave is breaking (see Fig. \ref{figure_momentum_flux}), this can shift the RMSE distribution.

\section{Conclusion}
\label{section_conclusion}

  In this study, an orographic source in a Lagrangian gravity-wave parameterization has been tested against wave-resolving simulations. The model, known as Multi-Scale Gravity-Wave Model or MS-GWaM \citep[e.g.][]{Muraschko2015, Boeloeni2016, Wilhelm2018, Wei2019}, solves prognostic equations that describe a superposition of quasilinear, locally monochromatic gravity-wave fields, based on WKB theory. More specifically, MS-GWaM uses ray-tracing techniques to integrate said equations along paths of constant wave-action density in a six-dimensional phase space. This method has certain advantages, such as the consideration of transience and horizontal wave propagation, without the risk of caustics due to crossing rays.

  So far, the orographic source has been implemented in a test version of the model. The latter is interactively coupled to the pseudo-incompressible flow solver PincFlow \citep{Rieper2013, Schmid2021}. A steady-state mode has been implemented, so that the importance of transient interaction between mean flow and gravity waves may be investigated.

  With these changes, the WKB model has been tested against wave-resolving simulations in an idealized setting. The latter features a two-dimensional, isothermal atmosphere with a monochromatic orography as a lower boundary. Various mountain heights, ranging from the linear to the nonlinear regime, have been used to simulate the generation of corresponding mountain waves over the period of one day. To ensure a gentle initialization, the orography is always linearly grown during the first three hours of the day. Wave reflections at the upper boundary are inhibited with a Rayleigh damping that increases exponentially with height \citep{Prusa2003}. In the wave-resolving simulations, this setup leads to an initially monochromatic and linear wave field that becomes nonlinear after some time, starting at a certain altitude that depends on the height of the mountains. Because of the horizontal periodicity of the domain, the flow is dominated by wave-mean flow interaction and blocking plays a less important role than in the situation considered by classic mountain wave parameterizations.

  In the WKB simulations, the transient model generally produced more accurate mean-wind profiles. The impact of transience becomes especially apparent in flows that remain linear at low altitudes. While the steady-state model predicts static instabilities as soon as a wave is generated that will eventually become unstable at a certain altitude, the transient model is able to mimic the slow propagation of wave energy that is seen in the wave-resolving simulations. In particular, the mean wind below the saturation level experiences a transient forcing that increases with altitude. As soon as the flow becomes nonlinear, the deviations from the wave-resolving simulations grow substantially larger. However, the mean wind forced by the transient model remains notably closer to the reference than the steady-state counterpart. This improvement is largest at high altitudes, where accounting for transience leads to the replication of a strong wind reversal. Steady-state parameterizations cannot induce this, due to the assumed formation of critical levels.

  The mean-wind RMSEs with respect to the wave-resolving reference are significantly smaller for the transient model. This improvement is especially pronounced in the linear regime, where the steady-state model produces errors more than twice as large.

  In summary, transience has been found to play an important role in the interaction between mean flow and mountain waves. By making use of single-column and steady-state approximations, classic mountain-wave parameterizations neglect the finite propagation speed of wave energy and exert a forcing on the mean flow that is unrealistic in the absence of wave breaking. Accounting for transient interaction allows for a more accurate simulation of the resolved dynamics, especially at high altitudes.

  It is to be emphasized that the present work only considered a highly idealized setting. Increasing the complexity will likely result in a different flow development. More realistic settings, such as isolated mountains in a domain with non-periodic lateral boundaries, as well as polychromatic and three-dimensional orographies will be the subject of future investigations. The goal of these will be to provide further physical and numerical insights for the main version of MS-GWaM \citep[e.g.][]{Boeloeni2021, Kim2021, Kim2024, Voelker2024}, in which an orographic source is currently being implemented.

\clearpage
\acknowledgments

F.J., R.C., F.L. and U.A. are grateful for support by Eric and Wendy Schmidt through the Schmidt Sciences VESRI ``DataWave'' project. U.A. thanks the German Research Foundation (DFG) for partial support through CRC 301 ``TPChange'' (Project No. 428312742, Projects B06 ``Impact of small-scale dynamics on UTLS transport and mixing'', B07 ``Impact of cirrus clouds on tropopause structure'' and Z03 ``Joint model development and modelling synthesis''). U.A. and G.S.V. thank the German Research Foundation (DFG) for partial support through the CRC 181 ``Energy transfers in Atmosphere and Ocean'' (Project No. 274762653, Projects W01 ``Gravity-wave parameterization for the atmosphere'' and S02 ``Improved Parameterizations and Numerics in Climate Models'').

%
%
\datastatement

The code and data described in this study are available on request.

%






%



\appendix[A]
\appendixtitle{Transition to terrain-following coordinates}

  Since the lower boundary at which the orographic source is implemented is the large-scale part of the surface topography, it is convenient to solve the equations in terrain-following coordinates, where the vertical origin is exactly that boundary. The coordinate system that is used in the WKB simulations is defined by
  \begin{align}
    \lef(\hat{x}, \hat{y}, \hat{z}\rig) & = \lef(x, y, L_z \frac{z - h_\mathrm{m}}{L_z - h_\mathrm{m}}\rig),
  \end{align}
  where $L_z$ is the vertical extent of the domain \citep{GalChen1975}. For an arbitrary field $\psi$, spatial derivatives in the Cartesian system relate to those in the terrain-following system via
  \begin{align}
    \pdx{\psi} & = \frac{1}{J} \lef(\frac{\partial J \psi}{\partial \hat{x}} + \frac{\partial J G^{1 3} \psi}{\partial \hat{z}}\rig),\\
    \pdy{\psi} & = \frac{1}{J} \lef(\frac{\partial J \psi}{\partial \hat{y}} + \frac{\partial J G^{2 3} \psi}{\partial \hat{z}}\rig),\\
    \pdz{\psi} & = \frac{1}{J} \frac{\partial \psi}{\partial \hat{z}},
  \end{align}
  where
  \begin{align}
    J = \frac{L_z - h_\mathrm{m}}{L_z}
  \end{align}
  is the Jacobian of the transformation and
  \begin{align}
    G^{1 3} & = \pdx{h_\mathrm{m}} \frac{\hat{z} - L_z}{L_z - h_\mathrm{m}}, & G^{2 3} & = \pdy{h_\mathrm{m}} \frac{\hat{z} - L_z}{L_z - h_\mathrm{m}}
  \end{align}
  are elements of the metric tensor \citep{GalChen1975, Clark1977}.

  With these relations, one can directly derive the prognostic equations in the transformed system. The right-hand side of \eqref{ray_equations_frequency} remains unchanged, as it only depends on a temporal derivative. However, the operator on the left-hand side of the equation is expressed differently, yielding
  \begin{align}
    \label{ray_equations_frequency_tfc}
    \pdt{\omega} + \frac{c_{\mathrm{g} x}}{J} \lef(\frac{\partial J \omega}{\partial \hat{x}} + \frac{\partial J G^{1 3} \omega}{\partial \hat{z}}\rig) + \frac{c_{\mathrm{g} y}}{J} \lef(\frac{\partial J \omega}{\partial \hat{y}} + \frac{\partial J G^{2 3} \omega}{\partial \hat{z}}\rig) + \frac{c_{\mathrm{g} z}}{J} \frac{\partial \omega}{\partial \hat{z}} = \pdt{\Omega}.
  \end{align}
  The prognostic equations for the wavenumbers also contain spatial derivatives on their right-hand sides, so that they change to
  \begin{align}
    \label{ray_equations_wavenumbers_k_tfc}
    \pdt{k} + \frac{c_{\mathrm{g} x}}{J} \lef(\frac{\partial J k}{\partial \hat{x}} + \frac{\partial J G^{1 3} k}{\partial \hat{z}}\rig) + \frac{c_{\mathrm{g} y}}{J} \lef(\frac{\partial J k}{\partial \hat{y}} + \frac{\partial J G^{2 3} k}{\partial \hat{z}}\rig) + \frac{c_{\mathrm{g} z}}{J} \frac{\partial k}{\partial \hat{z}} & = - \frac{1}{J} \lef(\frac{\partial J \Omega}{\partial \hat{x}} + \frac{\partial J G^{1 3} \Omega}{\partial \hat{z}}\rig),\\
    \label{ray_equations_wavenumbers_l_tfc}
    \pdt{l} + \frac{c_{\mathrm{g} x}}{J} \lef(\frac{\partial J l}{\partial \hat{x}} + \frac{\partial J G^{1 3} l}{\partial \hat{z}}\rig) + \frac{c_{\mathrm{g} y}}{J} \lef(\frac{\partial J l}{\partial \hat{y}} + \frac{\partial J G^{2 3} l}{\partial \hat{z}}\rig) + \frac{c_{\mathrm{g} z}}{J} \frac{\partial l}{\partial \hat{z}} & = - \frac{1}{J} \lef(\frac{\partial J \Omega}{\partial \hat{y}} + \frac{\partial J G^{2 3} \Omega}{\partial \hat{z}}\rig),\\
    \label{ray_equations_wavenumbers_m_tfc}
    \pdt{m} + \frac{c_{\mathrm{g} x}}{J} \lef(\frac{\partial J m}{\partial \hat{x}} + \frac{\partial J G^{1 3} m}{\partial \hat{z}}\rig) + \frac{c_{\mathrm{g} y}}{J} \lef(\frac{\partial J m}{\partial \hat{y}} + \frac{\partial J G^{2 3} m}{\partial \hat{z}}\rig) + \frac{c_{\mathrm{g} z}}{J} \frac{\partial m}{\partial \hat{z}} & = - \frac{1}{J} \frac{\partial \Omega}{\partial \hat{z}}.
  \end{align}
  On the other hand, the position tendencies \eqref{ray_equations_position} do not need to be changed. While one could certainly track the wave action in terrain-following coordinates, this is not necessary. Regardless of the coordinate system in which the rays are defined, evolving them requires the computation of large-scale quantities at their positions. By defining them in Cartesian coordinates, additional metric terms are avoided. The divergence on the left-hand side of the phase-space-wave-action-density equation \eqref{phase_space_wave_action_conservation} is replaced by its terrain-following counterpart, yielding
  \begin{align}
    \label{phase_space_wave_action_conservation_tfc}
    \pdt{\mathcal{N}} + \frac{c_{\mathrm{g} x}}{J} \lef(\frac{\partial J \mathcal{N}}{\partial \hat{x}} + \frac{\partial J G^{1 3} \mathcal{N}}{\partial \hat{z}}\rig) + \frac{c_{\mathrm{g} y}}{J} \lef(\frac{\partial J \mathcal{N}}{\partial \hat{y}} + \frac{\partial J G^{2 3} \mathcal{N}}{\partial \hat{z}}\rig) + \frac{c_{\mathrm{g} z}}{J} \frac{\partial \mathcal{N}}{\partial \hat{z}} + \dot{\vec{k}} \cdot \gra_\vec{k} \mathcal{N} = \sum\limits_s \mathcal{S}_s.
  \end{align}
  Note that the numerical implementation of the equations is unaffected by all left-hand-side changes, since the model integrates along paths defined by the phase-space velocities.

  The mean-flow tendencies \eqref{mean_flow_impact} are modified analogously, resulting in
    \begin{align}
      \lef(\pdt{u_\mathrm{m}}\rig)_\mathrm{w} & = - \frac{1}{J \bar{\rho}} \lef\{\frac{\partial}{\partial \hat{x}} \lef[J \int \hat{c}_{\mathrm{g} x} k \mathcal{N} \, \ddd V_\vec{k}\rig] + \frac{\partial}{\partial \hat{z}} \lef[J G^{1 3} \int \hat{c}_{\mathrm{g} x} k \mathcal{N} \, \ddd V_\vec{k}\rig]\rig.\notag\\
      & \qquad \qquad + \frac{\partial}{\partial \hat{y}} \lef[J \int \hat{c}_{\mathrm{g} y} k \mathcal{N} \, \ddd V_\vec{k}\rig] + \frac{\partial}{\partial \hat{z}} \lef[J G^{2 3} \int \hat{c}_{\mathrm{g} y} k \mathcal{N} \, \ddd V_\vec{k}\rig]\notag\\
      \label{mean_flow_impact_u}
      & \qquad \qquad + \lef.\frac{\partial}{\partial \hat{z}} \int c_{\mathrm{g} z} k \mathcal{N} \, \ddd V_\vec{k}\rig\},\\
      \lef(\pdt{\upsilon_\mathrm{m}}\rig)_\mathrm{w} & = - \frac{1}{J \bar{\rho}} \lef\{\frac{\partial}{\partial \hat{x}} \lef[J \int \hat{c}_{\mathrm{g} x} l \mathcal{N} \, \ddd V_\vec{k}\rig] + \frac{\partial}{\partial \hat{z}} \lef[J G^{1 3} \int \hat{c}_{\mathrm{g} x} l \mathcal{N} \, \ddd V_\vec{k}\rig]\rig.\notag\\
      & \qquad \qquad + \frac{\partial}{\partial \hat{y}} \lef[J \int \hat{c}_{\mathrm{g} y} l \mathcal{N} \, \ddd V_\vec{k}\rig] + \frac{\partial}{\partial \hat{z}} \lef[J G^{2 3} \int \hat{c}_{\mathrm{g} y} l \mathcal{N} \, \ddd V_\vec{k}\rig]\notag\\
      \label{mean_flow_impact_v}
      & \qquad \qquad + \lef.\frac{\partial}{\partial \hat{z}} \int c_{\mathrm{g} z} l \mathcal{N} \, \ddd V_\vec{k}\rig\}.
    \end{align}
    An additional mean-flow tendency is needed because the large-scale-flow solver predicts the terrain-following wind
    \begin{align}
      \hat{w}_\mathrm{m} = G^{1 3} u_\mathrm{m} + G^{2 3} \upsilon_\mathrm{m} + \frac{1}{J} w_\mathrm{m},
    \end{align}
    where $w_\mathrm{m}$ is a mean vertical wind that does not interact with the gravity waves considered in the present theory. The waves' impact on the terrain-following wind is given by
    \begin{align}
      \label{mean_flow_impact_w}
      \lef(\pdt{\hat{w}_\mathrm{m}}\rig)_\mathrm{w} = G^{1 3} \lef(\pdt{u_\mathrm{m}}\rig)_\mathrm{w} + G^{2 3} \lef(\pdt{\upsilon_\mathrm{m}}\rig)_\mathrm{w}.
    \end{align}

\appendix[B]
\appendixtitle{Discretization}

  \subsection{Transient model}
  \label{subsection_transient_model}

    The discretization of the ray equations requires the introduction of so-called ray volumes. These are initialized at every point in phase space where $\mathcal{N} \neq 0$ \citep{Muraschko2015}. The properties of a ray volume have a temporal index, three indices for physical space and a ray index $\beta$, with the range of the latter being the number of ray volumes existing at the considered point in time and physical space. The mean flow is computed on a staggered C-grid \citep{Arakawa1977} with cell centers defined by
    \begin{align}
      \hat{x}_i & = - \frac{L_x}{2} + \lef(i - 1\rig) \Delta \hat{x} + \frac{\Delta \hat{x}}{2}, & i & = 1, ..., N_x,\\
      \hat{y}_j & = - \frac{L_y}{2} + \lef(j - 1\rig) \Delta \hat{y} + \frac{\Delta \hat{y}}{2}, & j & = 1, ..., N_y,\\
      \hat{z}_k & = \lef(k - 1\rig) \Delta \hat{z} + \frac{\Delta \hat{z}}{2}, & k & = 1, ..., N_z.
    \end{align}
    For the sake of readability, all unshifted spatial indices will be omitted in the following.

    A ray volume located at $\lef(\hat{x}, \hat{y}, \hat{z}\rig)$ at the time $t^n$ is characterized by
    \begin{align}
      \vec{R}_\beta^n & = \lef(\vec{x}_\beta^n, \Delta \vec{x}_\beta^n, \vec{k}_\beta^n, \Delta \vec{k}_\beta^n, \hat{\omega}_\beta^n, \mathcal{N}_\beta^n\rig),
    \end{align}
    where $\lef(\vec{x}_\beta^n, \Delta \vec{x}_\beta^n, \vec{k}_\beta^n, \Delta \vec{k}_\beta^n\rig)$ denote the volume's location and extent in phase space, $\hat{\omega}_\beta^n$ is the intrinsic frequency and $\mathcal{N}_\beta^n$ the phase-space wave-action density. This means that in addition to the equations considered so far, prognostic equations for the extent of ray volumes are needed. The ray-volume surfaces $\Delta x_\beta^n \Delta k_\beta^n$, $\Delta y_\beta^n \Delta l_\beta^n$ and $\Delta z_\beta^n \Delta m_\beta^n$ are conserved in phase space \citep{Muraschko2015}. Thus, it is sufficient to predict the extent in physical space. From \eqref{ray_equations_position}, one can infer that
    \begin{align}
      \label{physical_extent_x}
      \ddt{\Delta x_\beta} & = \ddt{x_{\beta, +}} - \ddt{x_{\beta, -}} \approx \til{u}_{\mathrm{m}, \beta, +} - \til{u}_{\mathrm{m}, \beta, -},\\
      \label{physical_extent_y}
      \ddt{\Delta y_\beta} & = \ddt{y_{\beta, +}} - \ddt{y_{\beta, -}} \approx \til{\upsilon}_{\mathrm{m}, \beta, +} - \til{\upsilon}_{\mathrm{m}, \beta, -},\\
      \label{physical_extent_z}
      \ddt{\Delta z_\beta} & = \ddt{z_{\beta, +}} - \ddt{z_{\beta, -}} \approx \til{c}_{\mathrm{g} z, \beta, +} - \til{c}_{\mathrm{g} z, \beta, -},
    \end{align}
    where $\til{u}_{\mathrm{m}, \beta, \pm}$ is the linear interpolation of $u_\mathrm{m}$ to $x_{\beta, \pm} = x_\beta \pm \Delta x_\beta / 2$ and $\til{\upsilon}_{\mathrm{m}, \beta, \pm}$ is the equivalent for $\upsilon_\mathrm{m}$ in $y$-direction. Note that the intrinsic horizontal group velocities do not contribute to the evolution of $\Delta x_\beta$ and $\Delta y_\beta$, because their spatial dependence within a ray volume is purely vertical. In contrast, $\Delta z_\beta$ is affected by the vertical gradient of $N^2$, both implicitly and explicitly. This is accounted for in the computation of $\til{c}_{\mathrm{g} z, \beta, \pm}$, where the intrinsic frequency and squared buoyancy frequency are linearly interpolated to $z_{\beta, \pm} = z_\beta \pm \Delta z_\beta / 2$.

    The discretization of \eqref{ray_equations_wavenumbers}, \eqref{ray_equations_position} and \eqref{physical_extent_x}-\eqref{physical_extent_z} is performed with a third-order Runge-Kutta scheme \citep{Williamson1980} in time and second-order centered differences \citep[e.g.][p. 27]{Durran2010} in physical space. In the computation of the right-hand-side terms, all properties of the mean flow (such as $N^2$, $\vec{u}_\mathrm{m}^n$ and gradients thereof) are linearly interpolated to the physical ray-volume position. Note that \eqref{ray_equations_frequency} does not need to be integrated, since the intrinsic frequency can be recomputed from the updated wavenumbers and interpolated buoyancy frequency, with the frequency branch $\sigma \in \lef\{1, - 1\rig\}$ being a model parameter. In the phase-space-wave-action-density equation \eqref{phase_space_wave_action_conservation}, the dissipation term $\mathcal{S}_0$ is integrated with an explicit Euler step, whereas an implicit substep is performed at each Runge-Kutta stage for the Rayleigh damping term $\mathcal{S}_1$. The integrands in the discretized version of \eqref{diffusion_coefficient} are supplemented by the factor
    \begin{align}
      \label{diffusion_coefficient_factor}
      S_\beta^n = \max \lef(1, \frac{\Delta x_\beta^n}{\Delta \hat{x}}\rig) \max \lef(1, \frac{\Delta y_\beta^n}{\Delta \hat{y}}\rig) \max \lef(1, \frac{\Delta z_\beta^n}{J \Delta \hat{z}}\rig),
    \end{align}
    so that the contribution of each ray volume is weighted by the maximum grid cell fraction it can cover, regardless of whether it is partially outside of the local grid cell \citep{Boeloeni2016, Wei2019}.

    At the end of the third Runge-Kutta stage, the mean-flow impact is computed from the discretized version of \eqref{mean_flow_impact}. Therein, the contribution from each volume that is at least partially within $\lef(\hat{x} \pm \Delta \hat{x} / 2, \hat{y} \pm \Delta \hat{y} / 2, h_\mathrm{m} + J \hat{z} \pm J \Delta \hat{z} / 2\rig)^\mathrm{T}$ is weighted by the exact grid cell fraction it covers (note that this includes ray volumes that are assigned to adjacent grid cells). Before the tendencies are added to the evolution of the resolved flow, they are smoothed with a second-order Shapiro filter \citep{Shapiro1970}. This is done to remove small-scale features that may occur due to a coarse ray-volume distribution \citep{Muraschko2015}.

  \subsection{Boundaries}
  \label{subsection_boundaries}

    To limit the propagation of wave action to a constrained domain, boundary conditions are implemented in all spatial directions. In the horizontal, periodicity implies that ray volumes leaving the domain on one side reenter it from the opposite side. On the other hand, ray volumes that propagate beyond the upper or lower boundary should no longer affect the flow, which is guaranteed by setting their phase-space wave-action densities to zero, i.e.
    \begin{align}
      \mathcal{N}_{k = - 1, \beta}^n & = 0, & \mathcal{N}_{k = N_z + 1, \beta}^n & = 0.
    \end{align}

    In addition, the orographic source is implemented as a lower-boundary condition. This is done by launching correspondingly defined ray volumes at the first level below the ground. Note that ray volumes at the second level below the ground are subject to the boundary condition mentioned above. The number of ray volumes launched initially at a particular grid point is equal to the number of modes in the local spectrum of the orography's deviation from its background $h_\mathrm{m}$. The intrinsic frequency of a ray volume at the lower boundary is thus computed from the wavenumbers of the corresponding mode. Since the wind that drives the mountain waves should be above the lower boundary, one has
    \begin{align}
      \label{intrinsic_frequency_at_source}
      \hat{\omega}_{k = 0, \beta}^n = \sigma \lef|- k_{h, \beta} u_{\mathrm{m}, k = 1}^n - l_{h, \beta} \upsilon_{\mathrm{m}, k = 1}^n\rig|.
    \end{align}
    By prescribing the sign of $\hat{\omega}_{k = 0, \beta}^n$, the signs of the $k_{k = 0, \beta}^n$ and $l_{k = 0, \beta}^n$ are constrained as well. Note that this is purely due to the stationarity of mountain waves. The wave field is always a superposition of both frequency branches, which are accounted for by taking the real part and choosing a fixed $\sigma$. However, $\omega_{k = 0, \beta}^n = 0$ implies
    \begin{align}
      \label{intrinsic_frequency_at_source_with_correct_wavenumbers}
      \hat{\omega}_{k = 0, \beta}^n = - k_{k = 0, \beta}^n u_{\mathrm{m}, k = 1}^n - l_{k = 0, \beta}^n \upsilon_{\mathrm{m}, k = 1}^n,
    \end{align}
    so that for a given $\vec{u}_{\mathrm{m}, k = 1}^n$, either $\lef(k_{h, \beta}, l_{h, \beta}\rig)^\mathrm{T}$ or $\lef(- k_{h, \beta}, - l_{h, \beta}\rig)^\mathrm{T}$ are consistent with the chosen branch. The correct choice is ensured with
    \begin{align}
      \label{zonal_wavenumber_at_source}
      k_{k = 0, \beta}^n & = \sigma \sgn \lef(- k_{h, \beta} u_{\mathrm{m}, k = 1}^n - l_{h, \beta} \upsilon_{\mathrm{m}, k = 1}^n\rig) k_{h, \beta},\\
      \label{meridional_wavenumber_at_source}
      l_{k = 0, \beta}^n & = \sigma \sgn \lef(- k_{h, \beta} u_{\mathrm{m}, k = 1}^n - l_{h, \beta} \upsilon_{\mathrm{m}, k = 1}^n\rig) l_{h, \beta},
    \end{align}
    as can be checked by inserting \eqref{zonal_wavenumber_at_source}-\eqref{meridional_wavenumber_at_source} in \eqref{intrinsic_frequency_at_source_with_correct_wavenumbers} and comparing with \eqref{intrinsic_frequency_at_source}. The vertical wavenumber is given by
    \begin{align}
      m_{k = 0, \beta}^n = - \sigma \sqrt{\frac{\lef[\lef(k_{k = 0, \beta}^n\rig)^2 + \lef(l_{k = 0, \beta}^n\rig)^2\rig] \lef[N_{k = 1}^2 - \lef(\hat{\omega}_{k = 0, \beta}^n\rig)^2\rig]}{\lef(\hat{\omega}_{k = 0, \beta}^n\rig)^2}}.
    \end{align}

    To determine the phase-space wave-action density, the size of the ray volume must be specified. For the spatial extents, the obvious choice is
    \begin{align}
      \Delta \vec{x}_{k = 0, \beta}^n = \lef(\Delta \hat{x}, \Delta \hat{y}, J \Delta \hat{z}\rig)^\mathrm{T}.
    \end{align}
    However, no such choice exists for the spectral extents. Instead, they are determined from the wavenumbers, following
    \begin{align}
      \Delta \vec{k}_{k = 0, \beta}^n & = \lef(r_k \lef|k_{k = 0, \beta}^n\rig|, r_l \lef|l_{k = 0, \beta}^n\rig|, r_m \lef|m_{k = 0, \beta}^n\rig|\rig)^\mathrm{T},
    \end{align}
    where $r_k$, $r_l$ and $r_m$ are tuning parameters. The phase-space wave-action density at the source level is then given by
    \begin{align}
      \label{phase_space_wave_action_density_at_source}
      \mathcal{N}_{k = 0, \beta}^n = \frac{\bar{\rho}_{k = 1}}{2} \frac{\hat{\omega}_{k = 0, \beta}^n \lef|\vec{k}_{k = 0, \beta}^n\rig|^2}{\lef(k_{k = 0, \beta}^n\rig)^2 + \lef(l_{k = 0, \beta}^n\rig)^2} \frac{\lef|h_{\mathrm{w}, \beta}\rig|^2}{\Delta k_{k = 0, \beta}^n \Delta l_{k = 0, \beta}^n \Delta m_{k = 0, \beta}^n}.
    \end{align}

    \begin{figure*}
      \includegraphics[width = \textwidth]{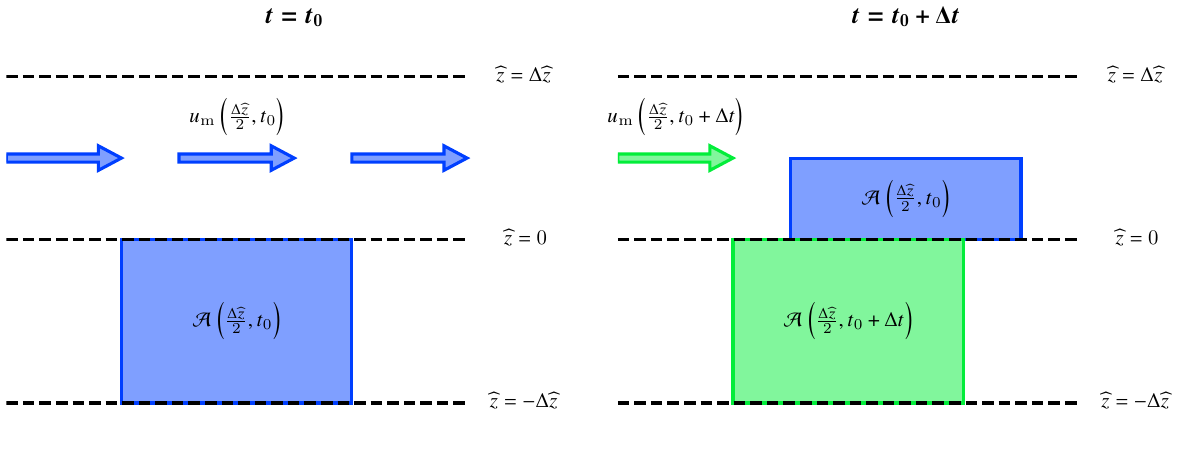}
      \caption{Schematic of the ray-volume launch algorithm. At $t = t_0$, a volume is launched with the wave-action density that is generated by the resolved wind crossing the unresolved mountains. One time step later, half of the volume has not yet passed the lower boundary and is therefore discarded. Meanwhile, a new volume is launched, according to the possibly changed resolved wind (inspired by Fig. 4 in \citet{Boeloeni2021}).}
      \label{figure_ray_volume_launch_algorithm}
    \end{figure*}

    Given the constrained properties of ray volumes at the lower boundary, it must be decided when they are to be launched. Simply adding new ray volumes at every Runge-Kutta stage may lead to unphysical behavior when the vertical group velocity is small. Instead, the model should check at every grid point $\lef(i, j, 0\rig)$ and for every mode $\beta$ whether there is still a ray volume with nonzero wave-action density before launching a new one. More specifically, three cases are distinguished.
    \begin{enumerate}
      \item There is no ray volume with nonzero wave-action density. A new ray volume is launched.
      \item There is a ray volume with nonzero wave-action density, which has partially passed the lower boundary. It is clipped and the part below the lower boundary is discarded before a new ray volume is launched.
      \item There is a ray volume with nonzero wave-action density, which has not yet crossed the lower boundary. It is replaced with a new one.
    \end{enumerate}
    Note that this is similar to the launch algorithm used by \citet{Boeloeni2021} for a simple representation of non-orographic gravity-wave sources. However, the present algorithm differs in that it allows for refraction immediately after the launch. Thus, both the shape and propagation speed of new ray volumes may change as soon as they start moving. This means it is technically possible for a newly launched ray volume to propagate beyond the first model level within one Runge-Kutta step. In practice, this does not happen because the time step of the model is constrained by a CFL condition with respect to the group velocity. An illustration of the algorithm is depicted in Fig. \ref{figure_ray_volume_launch_algorithm}.

  \subsection{Steady-state model}
  \label{subsection_steady_state_model}

    The numerical implementation of the model as described so far can be reduced to that of the steady-state model discussed in section \ref{section_theory}\ref{subsection_steady_state_theory}. Tracking ray volumes is no longer necessary due to the instantaneous distribution of wave action. The orographic source remains effectively unchanged, since the continuous launching of ray volumes is reduced to a regular update of the wave properties at the lower boundary (this is comparable to the third case described at the end of the previous section). From these, the wave field at higher levels can be determined, following the constraints of the steady-state theory.

    The horizontal wavenumbers are prescribed by the orographic source at the lowest level, whereas the intrinsic frequencies and vertical wavenumbers are determined from \eqref{steady_state_intrinsic_frequency} and \eqref{steady_state_vertical_wavenumber}, respectively. In the wave-action density equation \eqref{wave_action_conservation_steady_state}, the Rayleigh damping term is integrated with implicit Euler substeps. At each level, the dissipation term is integrated explicitly over a pseudo-time step $J \Delta \hat{z} / c_{\mathrm{g} z, \alpha}^n$, before the wave-action densities at the next level are determined. The turbulent viscosity and diffusivity \eqref{diffusion_coefficient_steady_state} is computed from all spectral modes. Critical and reflecting levels are accounted for by simply setting the wave-action density of the affected mode to zero when the local intrinsic frequency crosses the respective thresholds. Finally, the mean flow impact is obtained from the centered-differences discretization of \eqref{mean_flow_impact_steady_state}.

\appendix[C]
\appendixtitle{Shifting}

  The model stores ray volumes in a grid-cell-specific (four-dimensional) array. When the center of a ray volume moves beyond the grid cell it is currently assigned to, its position in the array is shifted accordingly. This algorithm also enforces the periodic boundary conditions in the horizontal and removes ray volumes that have left the domain in the vertical direction. Because it is performed only once per time step (before the mean-flow impact is computed), it cannot track ray volumes that move more quickly than $\lef(\Delta t\rig)^{- 1} \lef(\Delta \hat{x}, \Delta \hat{y}, J \Delta \hat{z}\rig)^\mathrm{T}$. For this reason, the time step is constrained by a CFL criterion with respect to the group velocity.

\appendix[D]
\appendixtitle{Splitting}

  \begin{figure}
    \includegraphics[width = \columnwidth]{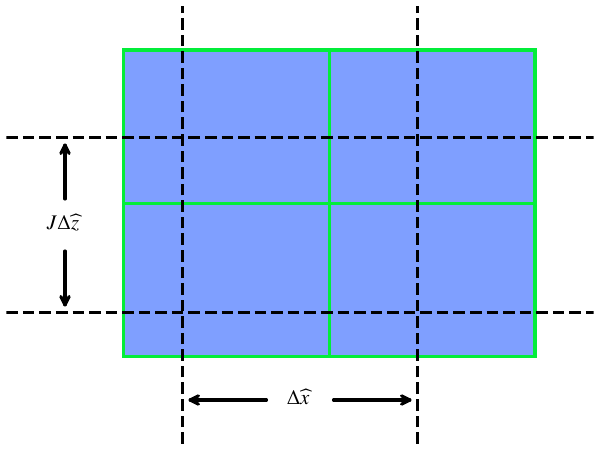}
    \caption{Schematic of a ray volume that is split in two dimensions.}
    \label{figure_ray_volume_split_algorithm}
  \end{figure}

  When a ray volume grows larger than the corresponding grid cell, assigning it to a single grid point becomes less accurate. To avoid this, a ray volume is split at its center when its extent in any direction exceeds the local grid spacing at the end of a time step (before the mean-flow impact is computed). The two new ray volumes then only differ from the old one in terms of their physical position and extent along the axis perpendicular to the split. This operation is performed recursively, so that a ray volume that fulfills the splitting criterion in all three directions is divided into exactly eight smaller ones. Fig. \ref{figure_ray_volume_split_algorithm} shows how the algorithm works in two dimensions.

\appendix[E]
\appendixtitle{Merging}

  \begin{figure}
    \includegraphics[width = \columnwidth]{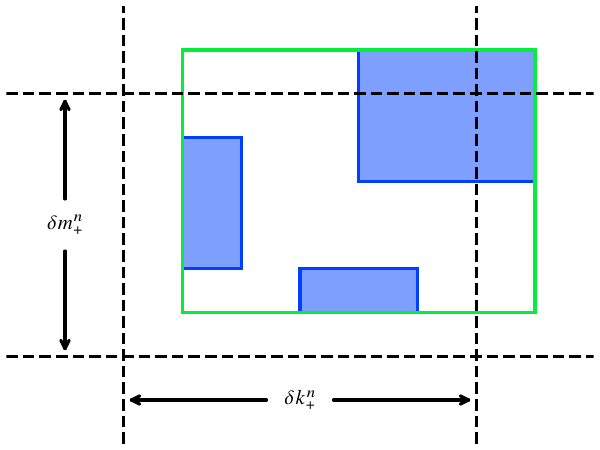}
    \caption{Schematic of three ray volumes that are merged in two spectral dimensions.}
    \label{figure_ray_volume_merge_algorithm}
  \end{figure}

  As a result of the splitting, the number of ray volumes may diverge to the point where the model becomes computationally unfeasible. To prevent this, the ray volumes in each grid cell are merged in spectral space when their number exceeds a critical value at the end of a time step (before the mean-flow impact is computed). For this purpose, a spectral range is determined from maxima and minima of all contributing wavenumbers. This range is split into three intervals for wavenumbers that are positive, negative and zero. The boundaries of these intervals are
  \begin{align}
    k_{\max, +}^n & = \max_\beta \lef\{k_\beta^n|k_\beta^n > 0\rig\}, & k_{\max, -}^n & = \max_\beta \lef\{k_\beta^n|k_\beta^n < 0\rig\},\\
    k_{\min, +}^n & = \min_\beta \lef\{k_\beta^n|k_\beta^n > 0\rig\}, & k_{\min, -}^n & = \min_\beta \lef\{k_\beta^n|k_\beta^n < 0\rig\},\\
    l_{\max, +}^n & = \max_\beta \lef\{l_\beta^n|l_\beta^n > 0\rig\}, & l_{\max, -}^n & = \max_\beta \lef\{l_\beta^n|l_\beta^n < 0\rig\},\\
    l_{\min, +}^n & = \min_\beta \lef\{l_\beta^n|l_\beta^n > 0\rig\}, & l_{\min, -}^n & = \min_\beta \lef\{l_\beta^n|l_\beta^n < 0\rig\},\\
    m_{\max, +}^n & = \max_\beta \lef\{m_\beta^n|m_\beta^n > 0\rig\}, & m_{\max, -}^n & = \max_\beta \lef\{m_\beta^n|m_\beta^n < 0\rig\},\\
    m_{\min, +}^n & = \min_\beta \lef\{m_\beta^n|m_\beta^n > 0\rig\}, & m_{\min, -}^n & = \min_\beta \lef\{m_\beta^n|m_\beta^n < 0\rig\}.
  \end{align}
  Logarithmic spacings are used to define the bins in which the ray volumes are to be merged. With the maximum number of ray volumes per grid cell being $N_k N_l N_m$, the spacings are given by
  \begin{align}
    \delta k_+^n & = \ln \lef(\frac{k_{\max, +}^n}{k_{\min, +}^n}\rig) \lef(\frac{N_k}{2} - 1\rig)^{- 1}, & \delta k_-^n & = \ln \lef(\frac{k_{\max, -}^n}{k_{\min, -}^n}\rig) \lef(\frac{N_k}{2} - 1\rig)^{- 1},\\
    \delta l_+^n & = \ln \lef(\frac{l_{\max, +}^n}{l_{\min, +}^n}\rig) \lef(\frac{N_l}{2} - 1\rig)^{- 1}, & \delta l_-^n & = \ln \lef(\frac{l_{\max, -}^n}{l_{\min, -}^n}\rig) \lef(\frac{N_l}{2} - 1\rig)^{- 1},\\
    \delta m_+^n & = \ln \lef(\frac{m_{\max, +}^n}{m_{\min, +}^n}\rig) \lef(\frac{N_m}{2} - 1\rig)^{- 1}, & \delta m_-^n & = \ln \lef(\frac{m_{\max, -}^n}{m_{\min, -}^n}\rig) \lef(\frac{N_m}{2} - 1\rig)^{- 1}.
  \end{align}

  All ray volumes in the considered cell are then binned according to their respective wavenumbers. In each bin $\gamma$, a new location and extent in phase space is determined from the maxima $\lef(\vec{x}_{\max, \gamma}^n, \vec{k}_{\max, \gamma}^n\rig)$ and minima $\lef(\vec{x}_{\min, \gamma}^n, \vec{k}_{\min, \gamma}^n\rig)$ of the collected volumes, i.e.
  \begin{align}
    \vec{x}_\gamma^n & = \frac{1}{2} \lef(\vec{x}_{\max, \gamma}^n + \vec{x}_{\min, \gamma}^n\rig), & \Delta \vec{x}_\gamma^n & = \vec{x}_{\max, \gamma}^n - \vec{x}_{\min, \gamma}^n,\\
    \vec{k}_\gamma^n & = \frac{1}{2} \lef(\vec{k}_{\max, \gamma}^n + \vec{k}_{\min, \gamma}^n\rig), & \Delta \vec{k}_\gamma^n & = \vec{k}_{\max, \gamma}^n - \vec{k}_{\min, \gamma}^n.
  \end{align}
  The new intrinsic frequency can then be calculated from the dispersion relation \eqref{dispersion_relation}. To ensure wave-energy conservation, the new phase-space wave-action density is computed from the total energy $E_{\sum, \gamma}^n$ of all contributing ray volumes, following
  \begin{align}
    \mathcal{N}_\gamma^n & = \frac{E_{\sum, \gamma}^n}{\hat{\omega}_\gamma^n \Delta x_\gamma^n \Delta k_\gamma^n \Delta y_\gamma^n \Delta l_\gamma^n \Delta z_\gamma^n \Delta m_\gamma^n}.
  \end{align}
  A schematic of how the merging works in two spectral dimensions is shown in Fig. \ref{figure_ray_volume_merge_algorithm}.


\bibliographystyle{ametsocV6}
\bibliography{references}

\end{document}